\documentclass[prl,twocolumn,showpacs,preprintnumbers]{revtex4-1}%
\usepackage{graphicx}
\usepackage{epstopdf}
\usepackage{array,multirow,amsmath,amsfonts}
\usepackage{color}
\usepackage{amsmath}
\usepackage{amsfonts}
\usepackage{amssymb}%
\providecommand{\U}[1]{\protect\rule{.1in}{.1in}}

\newcolumntype{Y}{>{\centering\arraybackslash}X}
\begin{document}
\preprint{ }
\title{SrRuO$_3$-SrTiO$_3$ heterostructure as a possible platform for studying unconventional superconductivity in  Sr$_{2}$RuO$_{4}$}
\author{Bongjae Kim$^{1,2}$}
\email{bongjae.kim@kunsan.ac.kr}
\author{Sergii Khmelevskyi$^{3}$}
\author{Cesare Franchini$^{4,5}$}
\author{I. I. Mazin$^{6,7}$}
\author{Kyoo Kim$^{2,8,9}$}
\email{kyoo@kaeri.re.kr}
\affiliation{$^{1}$ Department of Physics, Kunsan National University, Gunsan, 54150, Korea}
\affiliation{$^{2}$ MPPHC-CPM, Max Planck POSTECH/Korea Research Initiative, Pohang 37673, Korea}
\affiliation{$^{3}$ Center for Computational Materials Science, Institute for Applied
Physics, Vienna University of Technology, Wiedner Hauptstrasse $8$ - $10$,
$1040$ Vienna, Austria}
\affiliation{$^{4}$ University of Vienna, Faculty of Physics and Center for Computational
Materials Science, Vienna A-1090, Austria}
\affiliation{$^{5}$ Dipartimento di Fisica e Astronomia, Universit\`{a} di Bologna, 40127
Bologna, Italy}
\affiliation{$^{6}$ Code 6393, Naval Research Laboratory, Washington, DC 20375, USA}
\affiliation{$^{7}$ Quantum Materials Center, George Mason
University, Fairfax, VA 22030, USA }
\affiliation{$^{8}$ Department of Physics, Pohang University of Science and Technology,
Pohang 37673, Korea}
\affiliation{$^{9}$ Korea Atomic Energy Research Institute (KAERI), 111 Daedeok-daero, Daejeon 34057, Korea}
\date[Dated: ]{\today}

\begin{abstract}
There is intense controversy around the unconventional superconductivity in
strontium ruthenate, where the various theoretical and experimental studies
suggest diverse and mutually exclusive pairing symmetries. Currently, the
investigation is solely focused on only one material, Sr$_{2}$RuO$_{4}$, and
the field suffers from the lack of comparison targets. Here, employing a
density functional theory based analysis, we show that the heterostructure
composed of SrRuO$_{3}$ and SrTiO$_{3}$ is endowed with all the key
characteristics of Sr$_{2}$RuO$_{4}$, and, in principle, can host
superconductivity. Furthermore, we show that competing magnetic phases and
associated frustration, naturally affecting the superconducting state, can be
tuned by epitaxial strain engineering. This system thus offers an excellent
platform for gaining more insight into superconductivity in ruthenates.

\end{abstract}
\maketitle



\emph{Introduction. }
Transition metal oxides have long been a fertile ground for novel physics.
In particular, unconventional superconductivity, found, for instance, in cuprates and ruthenates,
is one of the most intriguing discovery of recent decades, and the physics and mechanisms in these
oxides is still highly debated.
The latter compound is particularly intriguing; despite tens of
years of intensive studies, even a basic understanding of pairing symmetry is
still lacking~\cite{Mackenzie2017}, and new surprises are continuously coming
up. Arguably, one of the reasons the latter compound has been so difficult to
crack is that it is a one-of-a-kind material. As opposed to the Cu- and Fe-based
high-$T_{c}$  superconductors, which have multiple families with many members
each, Sr$_{2}$RuO$_{4}$ is a family of one, and no insightful comparison with
any sibling can be effected.

In fact, one of the biggest breakthroughs in the field, which led to dramatic
progress, occurred when experimentalists found a way to apply uniaxial strain
so as to induce a Lifshitz transition~\cite{Hicks2014,Steppke2017}. Finding
another material, affording even more flexibility in modifying Sr$_{2}%
$RuO$_{4}$ properties, including its Fermiology and magnetic response, would
open up a cornucopia of new experimental information and may lead to
extraordinary progress in understanding the superconductivity in Sr$_{2}%
$RuO$_{4}$.

This idea is not new. Burganov \emph{et al.} demonstrated that it is
possible to grow epitaxial Sr$_{2}$RuO$_{4}$ thin films on a SrTiO$_{3}$
substrate, with a lattice mismatch of 0.9{\%}~\cite{Burganov2016}. No
superconductivity was observed, which may be related to either too many
defects at the interface (and the authors were optimistic about reducing
their amount in the future), or due to a $\sqrt{2}\times\sqrt{2}$ reconstruction, also
present at the free-standing Sr$_{2}$RuO$_{4}$ surface. In particular, they
observed in angle-resolved photoemission spectroscopy a Lifshitz transition
not unlike the one seen under uniaxial strain, and speculated that approaching
the van Hove singularity brings about a strong enhancement of the uniform
magnetic susceptibility $via$ the Stoner mechanism and therefore,
strengthening of ferromagnetic (FM) spin fluctuations.

Very recently, such an enhancement has been observed in the
resistivity~\cite{Barber2018}, as well as in nuclear magnetic resonance (NMR)
experiments and density functional theory (DFT) calculations~\cite{Luo2019} for bulk Sr$_{2}$RuO$_{4}$
under uniaxial stress. Note that, generally speaking, FM fluctuations favor
triplet, and antiferromagnetic fluctuations favor singlet pairing. In the
unstrained Sr$_{2}$RuO$_{4}$, the spin fluctuation spectrum is well
documented~\cite{Steffens2019}, and it has been demonstrated that, barring
other triplet-favoring interactions, the antiferromagnetic fluctuations win
hands down. Yet, if pairing is indeed triplet, superconductivity which cannot
benefit directly from the increased density of states (DOS) at the van Hove singularity can be
boosted by these FM fluctuations, in accordance with the observed $T_{c}$
behavior~\cite{Steppke2017}.

Not unexpectedly, the work by Burganov \emph{et al.}~\cite{Burganov2016} has
triggered several works based on theoretical modelling employing tight-binding
models~\cite{Hsu2016,Liu2018}. Unfortunately, such models do not take into
account all the complexity of the epitaxial interface, band effects, and structural
delicacy, and an extension to other systems has not been made so far. Here, we
exploit the idea of heterostructures, employing first principles calculations,
which, differently from model-based approaches, have a capacity to capture
much of the aforementioned complexity.

Curiously, while the cuprates, which are already a rather rich family, have
been subjected to multiple suggestions and attempts to generate similar
materials, including engineered heterostructures\cite{Arita2007_1,Anisimov1999,Chaloupka2008,Hansmann2009,Watanabe2013,YKKim2014,Gaw2019,Isaacs2019,Li2019}%
, Sr$_{2}$RuO$_{4}$, which needs expansion onto other systems much more badly,
has hardly been discussed in this context. Yet, from the perspective of the
growth technique, epitaxy and heterostructuring are well-understood for
ruthenates, especially for SrRuO$_{3}$~\cite{Koster2012}. Particularly
promising would be a superlattice composed of a single layer of SrRuO$_{3}$
sandwiched between insulating layers, which is now experimentally accessible
with fine control~\cite{Gu2012,Boschker2019,Jeong2020}. It has the
same principal structural motif as Sr$_{2}$RuO$_{4}$, a two-dimensional square
RuO$_{2}$ lattice. As we show below, it shares with Sr$_{2}$RuO$_{4}$ key
features of the electronic structure and magnetic properties, but also has
interesting and promising distinctions.

In the following, we report a first principles computational study of the
electronic structure and magnetic properties of the SrRuO$_{3}$-SrTiO$_{3}$
(SRO-STO) heterostructure, grown on a STO substrate (we will also briefly
discuss other potential substrate ranges). The role of substrate is to fix the
lateral lattice dimensions, and, thus, to provide biaxial strain. We show that
the SRO-STO superlattice has all the key characteristics of Sr$_{2}$RuO$_{4}$,
and would broaden the exploratory potential dramatically in regard to
the superconductivity mechanism and pairing symmetry.

\begin{figure}[t]
\begin{center}
\includegraphics[width=85mm]{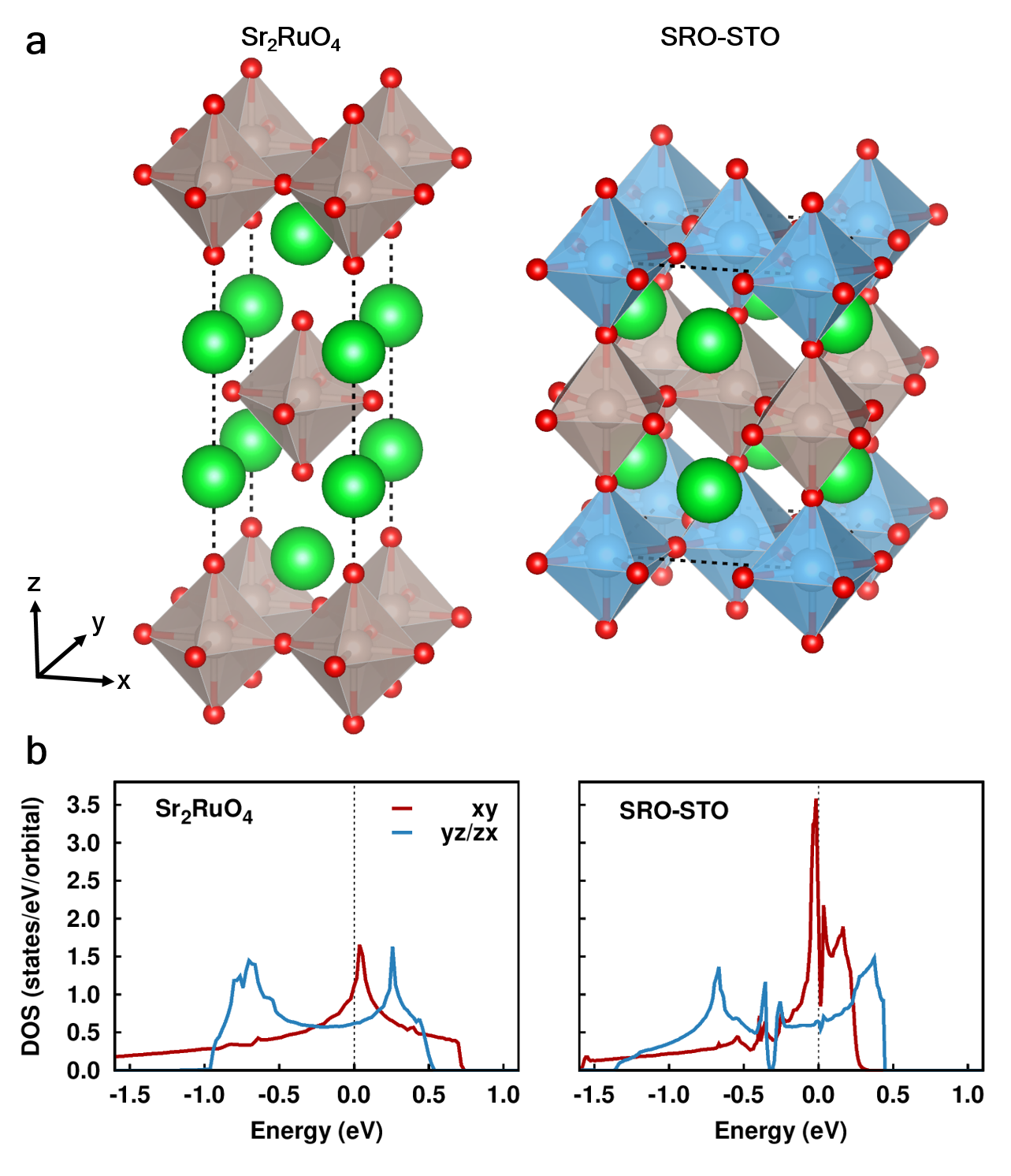}
\end{center}
\caption{(a) Crystal structure of Sr$_{2}$RuO$_{4}$ and SRO-STO. Gray and
blue octahedera denote RuO$_{6}$ and TiO$_{6}$ octahedra, and red and green
spheres represent oxygen and strontium atoms, respectively (b) Partial DOS of
Ru-$d$ orbitals for Sr$_{2}$RuO$_{4}$ and SRO-STO for ${xy}$ and ${yz/zx}%
$. }%
\label{fig1}%
\end{figure}

\emph{Result and discussion.}
The most important structural feature of Sr$_{2}$RuO$_{4}$ is the
two-dimensional (2D) RuO$_{2}$ layer, spaced between SrO layers, where the
corner-shared RuO$_{6}$ octahedra form a square net. This active layer can be
artificially constructed by sandwiching a SRO monolayer between STO blocks as
shown in Fig.~\ref{fig1} (a), where the out-of-plane Ru-O-Ru connectivity is
broken due to the interspaced STO layers. A small in-plane octahedral rotation
($\sim8.2^{\circ}$ for unstrained case) exists for this superlattice, which
produces a $\sqrt{2}\times\sqrt{2}$ supercell reconstruction in the
plane~\cite{Gu2012}. This reconstruction is not found for the pure bulk phase
of Sr$_{2}$RuO$_{4}$, but is observed at the surface, as well as in a Ca-doped
system~\cite{Matzdorf2000,Damascelli2000,Wang2004}, reflecting the tendency
for RuO$_{6}$ octahedra to rotate, if necessary to accommodate the geometrical
constraints. Note that even with this octahedral rotation pattern, surface
superconductivity has been observed in tunneling~\cite{Firmo2013}.

The structural similarity of the two systems is reflected in their electronic
structures. In Fig.~\ref{fig1} (b), we plotted the orbitally resolved partial
density of states (DOS) of Ru-$d$ for both systems. Due to an octahedral
crystal field splitting, the three $t_{2g}$ orbitals ($xy$, $yz$ and $zx$)
share four electrons, and are responsible for the low-energy physics.
The $xy$ orbital forms a quasi-2D band, and a nearly circular Fermi surface (FS),
usually labeled as $\gamma$, and the other orbitals form two quasi-1D bands,
which, upon intersection and re-hybridization, create two rounded-square
shaped Fermi surfaces ($\alpha$ and $\beta$). From the
partial DOS of SRO-STO, characteristic features observed in Sr$_{2}$RuO$_{4}$
are present as well. The peak of the $xy$ partial
DOS, corresponding to a 2D van Hove singularity (vHs), is located close to the
Fermi level, and its contribution at the Fermi energy is much larger than those of the
$yz/zx$ bands. Less dispersive $yz/zx$ orbitals have double-peak structures,
corresponding to 1D vHs, and have smaller contributions at the Fermi level.

In Fig. \ref{fig1} (b), one can notice that the $xy$ DOS peak in the
heterostructure is higher, and the bandwidth is smaller, than in Sr$_{2}%
$RuO$_{4}.$ This can be attributed to the octahedral rotations, which reduce the
effective $xy-xy$ hopping as $\cos^{2}\alpha,$ where $\alpha$ is the
rotational angle, while the $xz-xz$ and $yz-yz$ hopping is unaffected in the
same order in $\alpha.$ This fact is immediately relevant for one of the main
points of contention regarding superconductivity in Sr$_{2}$RuO$_{4}$: while
one school considers the $xy$ band to have strong pairing, with
superconductivity in $xz/yz$ being induced by an interband proximity
effect~\cite{Agterberg1997,Yanase2003}, another advocates the $xz/yz$ bands as
the \textquotedblleft active\textquotedblright\ subsystem, with $xy$ being
secondary~\cite{Raghu2010}. Yet others suggested that all bands contribute
roughly equally~\cite{Mazin1997,Mazin1999}. Enhancing selectively the DOS in
the $xy$ band only, through octahedral rotations (which, as we show later, can
be controlled by the substrate), provides a direct test of these alternatives.

Another issue of relevance is the proximity to magnetism. It is well known
that Sr$_{2}$RuO$_{4}$ is on the verge of an antiferromagnetic instability,
driven by the quasi-1D nesting in the $xz/yz$ bands. The corresponding spin
fluctuations favor $d$-wave pairing~\cite{Mazin1999,Steffens2019}. However,
there is also, albeit a much weaker, tendency to a FM instability, which favors
$p$-wave pairing. A dramatic increase in DOS brings the system much closer to
ferromagnetism, thus strongly pushing the system toward a triplet pairing.
Note that within the weak coupling limit, recent $T_{c}$  enhancement in
uniaxial strain experiment has been associated with odd- to even-parity
transition, and our study can offer one way to verify this
proposition~\cite{Steppke2017}.

With this in mind, we have looked at the magnetism of this system in more
detail. Note that there have been DFT-based studies on SRO-STO system both in
film and superlattices
structures~\cite{Jeong2020,Mahadevan2009,vAlves2012,Si2015}, and they
concluded the magnetic and electronic ground states of a single SRO layer to be
either FM metal or N\'{e}el-type antiferromagnetic insulator. This issue is
not also resolved as judged from the diverse experimental
reports~\cite{Gu2012,Boschker2019,Jeong2020,Chang2009}. The problem with these
studies is that they take the DFT ground state at its face value, forgetting
that it is, by nature, a mean-field approach, and, as such, liable to
overestimate the tendency to magnetism. Indeed, straight DFT calculations
predict Sr$_{2}$RuO$_{4}$ to be spin density wave (SDW) type antiferromagnetic
(not a N\'{e}el-type, see below), and, with gradient corrections, even the FM
state has lower energy than the nonmagnetic one~\cite{BKim2017}. This tendency
can be corrected for, phenomenologically, by introducing the concept of a
fluctuations-renormalized Stoner factor~\cite{Larson2004,Ortenzi2012}, or
strongly remedied by switching to dynamical
mean field theory~\cite{Kugler2019}. Applying the same concept to Sr$_{2}%
$RuO$_{4}$ and SRO-STO, we observed that the experimentally observed Stoner
enhancement for the former is 7~\cite{Steffens2019}, thus the Stoner product
$IN(E_{F})\approx0.857$ [at the same time, renormalization at $\mathbf{q}=\mathbf{Q_{SDW}}$ is
about 30, implying that $I(Q_{SDW})\chi_{0}(Q_{SDW})\approx0.967$, where
$\chi_{0}$ is bare susceptibility]. Our calculations find that the partial
density of states of Ru orbitals on the Fermi level is about 50\% higher in
SRO-STO. This suggests that their $IN(E_{F})$ is larger than one and the
ground state should be FM. Fixed spin moment calculations confirm that in
order to stabilize the paramagnetic state one would need to reduce the Stoner
$I$ by 3.5 times, which is completely unphysical.

\begin{figure}[t]
\begin{center}
\includegraphics[width=85mm]{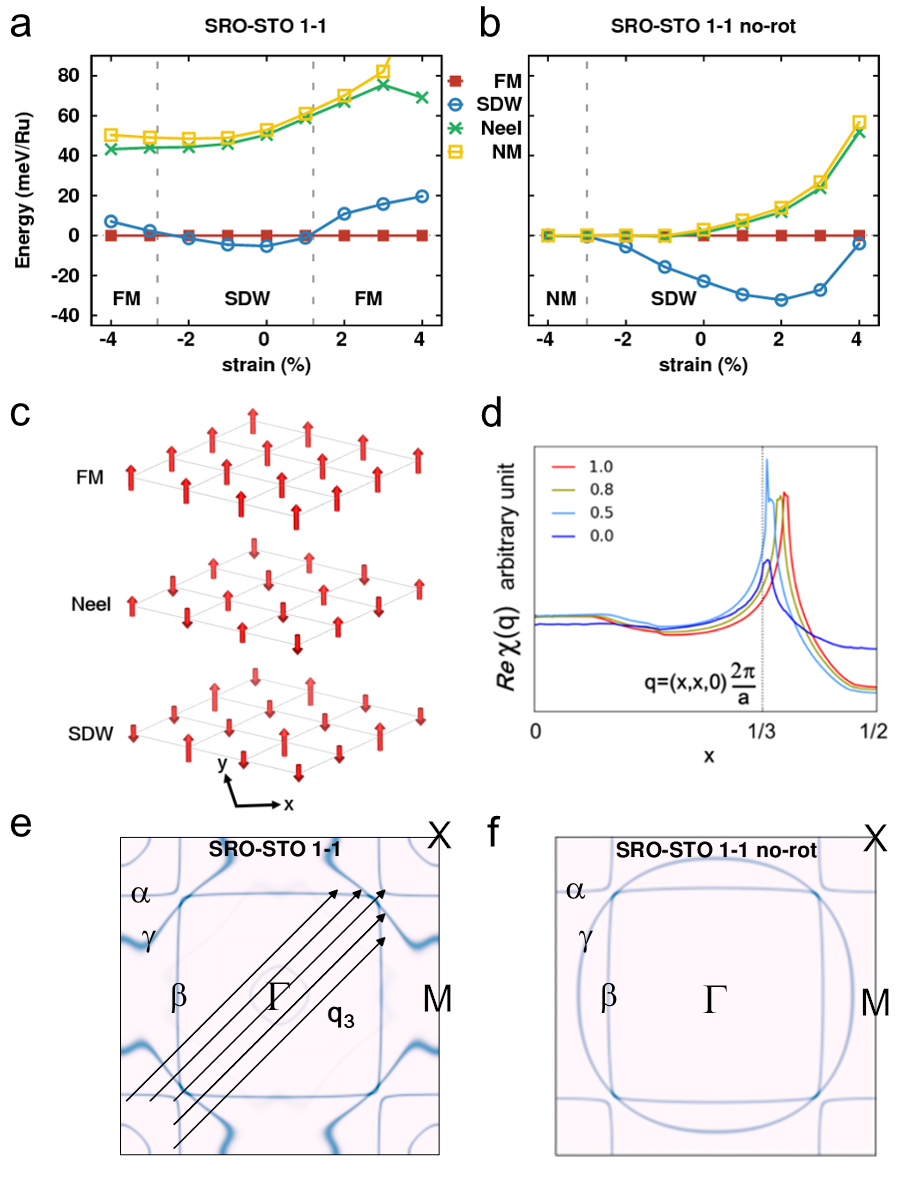}
\end{center}
\caption{The energy landscape as a
function of strain (a) with and (b) without octahedral rotation of RuO$_{6}$,
for three magnetic configurations schematically depicted in (c): FM
(ferromagnetic), N\'{e}el antiferromagnetic, and SDW ($\mathbf{q}%
=(1,1,0)\frac{2\pi}{3a}$) phases. (d) The change of Lindhard susceptibility
obtained from the unfolded band for Ru ${yz/zx}$ orbitals. Octahedral
rotations are linearly interpolated between the unrotated case (0.0) and the fully
rotated one (1.0). (e,f) The change of Fermi surfaces (e) with and (f) without
octahedral rotation. (d)-(f) are for the unstrained (0\%) structure. The
arrows in the Fermi surfaces indicate q$_{3}$ nesting vector which activates
both the ${xz}$ and ${yz}$ quasi-1D-Fermi surfaces. }%
\label{fig2}%
\end{figure}

As discussed below, the DOS at the Fermi level strongly depends on the
epitaxial strain, which offers a unique opportunity to tune the system all the
way from low DOS and low $T_{c}$  to higher DOS and high $T_{c}$, as in the
case of uniaxial strain~\cite{Hicks2014,Steppke2017}, and further to even
higher DOS, \emph{i.e.} the FM state. Furthermore, for a range of strains the FM
instability enters a fierce competition with the SDW one, which is even more
enhanced for the SRO-STO system (See Fig.~\ref{fig2}(a-c)). Bulk Sr$_{2}%
$RuO$_{4}$ is, both experimentally and theoretically, much closer to a SDW
instability than to a FM one. Strong magnetic fluctuations have been observed
by the neutron diffraction, which persists up to room
temperature~\cite{Sidis1999,Iida2011}. This had been predicted in DFT
studies~\cite{Mazin1999} and a SDW with $\mathbf{q}_{3}=(1,1,0)\frac{2\pi}%
{3a}$, a close commensurate approximation to the experimental peak in spin
susceptibility, was shown to be stable in calculations~\cite{BKim2017,Cobo2016}. On
the contrary, for the bulk material the FM state is barely stable in GGA, and
unstable in LDA. There is no magnetic frustration in the system, with the SDW
being a clear mean field ground state, suppressed by quantum fluctuations.

Interestingly, in SRO-STO, while both FM and SDW states gain stability,
eventually they become nearly degenerate (Fig.~\ref{fig2}(a) and (b)), adding a strong parametric
frustration to the system.
In a number of materials, most notably in Fe-based superconductors, such a strong magnetic frustration leads to
complete destruction of a long-range magnetic order and numerous interesting phenomena, such as nematic order,
and, arguably, superconductivity itself~\cite{Mazin2008,Glasbrenner2015}. It looks plausible that magnetic frustration in SRO-STO would prevent
the system from developing a static magnetic order, but would trigger strong spin ﬂuctuations, with all ramiﬁcations
for the superconductivity. While the jury is still out on whether spin-fluctuation mechanism is instrumental in cuprates,
Fe pnictides, or Sr$_2$RuO$_4$, it is considered by many to be a strong contender in all three cases (which, of course,
does not imply that the nature of superconductivity is the same in all three instances). Enhanced frustration of magnetism
in SRO-STO, in this context, looks especially interesting.

Previous computational studies of SRO-STO superlattice concentrated upon the
competition between FM and N\'{e}el orders~\cite{Mahadevan2009,Si2015},
ignoring the the SDW $\mathbf{q}_{3}=(1,1,0)\frac{2\pi}{3a}$ phase. Our
calculations show that, in spite of a dramatic stabilization of the FM phase,
the SDW still has a lower energy (Fig.~\ref{fig2}(a)) over a large range of
strains (the N\'{e}el order is not competitive at all). {The enhanced
competition between the two magnetic phases in SRO-STO is promoted by the octahedral
rotations, which are absent in Sr$_{2}$RuO$_{4}$. As shown in Fig.~\ref{fig2} (b),
the calculations with suppressed octahedral rotations favor the SDW tendency over FM.}
For strains between $\sim-2$\% and $\sim$1\%, the SDW has the lowest energy, but the energy
gain is minimal, $\sim$ 5 meV/Ru for zero strain case (the corresponding
value for Sr$_{2}$RuO$_{4}$ is $\sim$20 meV/Ru). For larger strains
($\lesssim-$2\% and $\gtrsim$1\%), the FM state becomes the most stable one.
Note that the critical strain for the transition from SDW to FM is within the accessibility
of current experimental techniques~\cite{Vailionis2011}, and even broader epitaxial strain
ranges could be reached in recent experiments~\cite{Schlom2007}. This suggests that the
SRO-STO system is a suitable platform for the control of different types of magnetic
interactions, and, by implication, for tuning the superconducting order parameter.
Note that, given that the fluctuation-induced suppression of magnetism may be,
and probably is, $\mathbf{q}$-dependent, it is hard to say which state has the
lowest free energy for which strain.

As can be seen in Fig.~\ref{fig2}(d), the real part of the charge susceptibility
in the SRO-STO system (see Supplementary Materials~\cite{suppl}) shows only a gradual
shift of the $\mathbf{q}_{3}$ peak upon octahedral rotation. This is consistent
with the fact that the $\alpha/\beta$ FSs remain quasi-one dimensional, with the
same $\mathbf{q}_{3}$ nesting vector, as indicated by arrows in Fig.~\ref{fig2} (e).
This lets us conclude that the
antiferromagnetic SDW fluctuations are highly robust against octahedral
rotation, and, consequently, against substrate engineering. However, the
two-dimensional $\gamma$ FS is highly sensitive to octahedral rotations,
which can be efficiently tuned by epitaxial strain. In
particular, $\gamma$ FS changes its shape with strain from a circle to a
rounded square, which is not related to the van Hove singularity often
discussed in strain studies of Sr$_{2}$RuO$_{4}$%
~\cite{Hicks2014,Steppke2017,Luo2019}, but is due to mixing between the $xy$
and $x^{2}-y^{2}$ orbitals of neighboring Ru atoms as shown in
Figs.~\ref{fig2}(e,f).
The hybridization between $xy$ and $x^{2}-y^{2}$ bands due to octahedral rotation is eventually
responsible for the additional FSs at around the $\Gamma$ and X point for SRO-STO (Fig.~\ref{fig2}(e))~\cite{suppl}.

\begin{figure}[ptb]
\par
\begin{center}
\includegraphics[width=85mm]{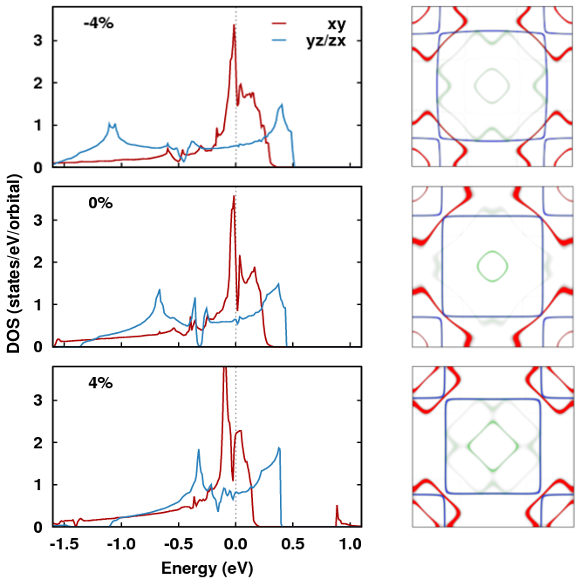}
\end{center}
\caption{Partial
DOSs of Ru $xy$ (red) and Ru $yz/zx$ (blue) bands for strains -4\%, 0\%, and
4\%. Corresponding Fermi surfaces including the $\Gamma$ point are shown in the right
column with the same color code. The contribution of other orbitals at Fermi
surface is is marked by the green color. }%
\label{fig3}%
\end{figure}

Changes of DOSs and FSs upon strain are shown in Fig.~\ref{fig3}. The width of
$yz/zx$ DOS becomes smaller as the tensile strain increases, while the $xy$
DOS is nearly unaffected except for a shift of the peak position toward
lower energies. Notably, the peak, mainly from the $xy$ orbital, which is
thought to be closely related with the $T_{c}$  enhancement of uniaxial strain
case, progressively moves down away from the Fermi level for tensile
strains while the $yz/zx$ DOS gradually increases at the Fermi level~\cite{comment,suppl}.
For Sr$_{2}$RuO$_{4}$, octahedral distortion shifts the peak down~\cite{Ko2007},
while in SRO-STO, a tensile strain, which generally relieves the distortion, has a
similar effect.

This selective evolution of the $xy$ and ${yz/zx}$ orbital upon strain
suggests that future experimental studies of the SRO-STO system can probe the
\textquotedblleft active\textquotedblright\ orbitals, important for
superconductivity~\cite{Agterberg1997,Yanase2003,Raghu2010}. Futhermore, the
heterostructural features of the SRO-STO system itself can restrict specific types
of scattering, and, within this structural setup, some superconducting
symmetries, such those with horizontal node lines, cannot survive. The DOS at the
Fermi level, which is tunable in our setup, is closely related with $T_{c}$
behavior for a singlet pairing, but not for a triplet one. Thus, the possibility of a
specific pairing such as $d_{x^{2}-y^{2}}$, which is promoted by the enhancement
of the DOS at the Fermi level, can be directly tested in the SRO-STO system.
{Being close relatives of the orphan superconductor Sr$_{2}$RuO$_{4}$,
SRO-STO superlattices offer a unique possibility to controllably tune the
Fermi surface topology, magnetic interactions and the position of the van Hove peak,
 thus providing new avenues to probe the paring symmetry
of the ruthenates.}

\emph{Conclusion.}
SRO-STO heterostructures present an electronic system in many aspects similar
to the intriguing putative triplet superconductor, Sr$_{2}$RuO$_{4},$ but also
distinctly different. The important common motif is the presence of the three
Fermi surfaces, representing three Ru orbitals, $xy,$ $yz$ and $zx,$ with the
first one exhibiting a van Hove singularity at or near the Fermi level.
Moreover, the exact topology of the Fermi surfaces can be modified within a
reasonable range by varying the substrate and thus exerting different
epitaxial strains. An important difference from Sr$_{2}$RuO$_{4}$ is that
SRO-STO appears, in general, more magnetic, and at the same time the tendency
to ferromagnetism is enhanced much more strongly than to the SDW. In principle,
three scenarios can be realized, all three interesting in their own way.
First, the enhanced tendency to SDW could lead to an actual instability.
Second, at least for some strains, the material may become FM. The third, and
arguably the most interesting, alternative is that the near-degeneracy (on a
mean field level) of the FM and the SDW state will lead to a very strong
parametric magnetic frustration that will preclude either static order, but
will greatly enhance both types of fluctuations. In this case, the system, if
manufactured cleanly enough, is likely to become superconducting, possibly in
the triplet channel (boosted by FM fluctuations), but the properties of this
superconducting state may be even more different from the bulk Sr$_{2}%
$RuO$_{4}$ than the critically strained samples~\cite{Steppke2017}. If the
symmetry of the superconducting state in Sr$_{2}$RuO$_{4}$ is indeed triplet
(which is, however, strongly questioned by recent NMR experiments~\cite{Pustogow2019,Ishida2019}), the
critical temperature of SRO-STO may be dramatically higher, because of stronger fluctuations,
than that of Sr$_{2}$RuO$_{4}$.


\bibliographystyle{apsrev4-1}
\bibliography{bibfile}

\begin{thebibliography}{60}%
\makeatletter
\providecommand \@ifxundefined [1]{%
 \@ifx{#1\undefined}
}%
\providecommand \@ifnum [1]{%
 \ifnum #1\expandafter \@firstoftwo
 \else \expandafter \@secondoftwo
 \fi
}%
\providecommand \@ifx [1]{%
 \ifx #1\expandafter \@firstoftwo
 \else \expandafter \@secondoftwo
 \fi
}%
\providecommand \natexlab [1]{#1}%
\providecommand \enquote  [1]{``#1''}%
\providecommand \bibnamefont  [1]{#1}%
\providecommand \bibfnamefont [1]{#1}%
\providecommand \citenamefont [1]{#1}%
\providecommand \href@noop [0]{\@secondoftwo}%
\providecommand \href [0]{\begingroup \@sanitize@url \@href}%
\providecommand \@href[1]{\@@startlink{#1}\@@href}%
\providecommand \@@href[1]{\endgroup#1\@@endlink}%
\providecommand \@sanitize@url [0]{\catcode `\\12\catcode `\$12\catcode
  `\&12\catcode `\#12\catcode `\^12\catcode `\_12\catcode `\%12\relax}%
\providecommand \@@startlink[1]{}%
\providecommand \@@endlink[0]{}%
\providecommand \url  [0]{\begingroup\@sanitize@url \@url }%
\providecommand \@url [1]{\endgroup\@href {#1}{\urlprefix }}%
\providecommand \urlprefix  [0]{URL }%
\providecommand \Eprint [0]{\href }%
\providecommand \doibase [0]{http://dx.doi.org/}%
\providecommand \selectlanguage [0]{\@gobble}%
\providecommand \bibinfo  [0]{\@secondoftwo}%
\providecommand \bibfield  [0]{\@secondoftwo}%
\providecommand \translation [1]{[#1]}%
\providecommand \BibitemOpen [0]{}%
\providecommand \bibitemStop [0]{}%
\providecommand \bibitemNoStop [0]{.\EOS\space}%
\providecommand \EOS [0]{\spacefactor3000\relax}%
\providecommand \BibitemShut  [1]{\csname bibitem#1\endcsname}%
\let\auto@bib@innerbib\@empty
\bibitem [{\citenamefont {Mackenzie}\ \emph {et~al.}(2017)\citenamefont
  {Mackenzie}, \citenamefont {Scaffidi}, \citenamefont {Hicks},\ and\
  \citenamefont {Maeno}}]{Mackenzie2017}%
  \BibitemOpen
  \bibfield  {author} {\bibinfo {author} {\bibfnamefont {A.~P.}\ \bibnamefont
  {Mackenzie}}, \bibinfo {author} {\bibfnamefont {T.}~\bibnamefont {Scaffidi}},
  \bibinfo {author} {\bibfnamefont {C.~W.}\ \bibnamefont {Hicks}}, \ and\
  \bibinfo {author} {\bibfnamefont {Y.}~\bibnamefont {Maeno}},\ }\href
  {https://www.nature.com/articles/s41535-017-0045-4} {\bibfield  {journal}
  {\bibinfo  {journal} {npj Quantum Materials}\ }\textbf {\bibinfo {volume}
  {2}},\ \bibinfo {pages} {40} (\bibinfo {year} {2017})}\BibitemShut {NoStop}%
\bibitem [{\citenamefont {Hicks}\ \emph {et~al.}(2014)\citenamefont {Hicks},
  \citenamefont {Brodsky}, \citenamefont {Yelland}, \citenamefont {Gibbs},
  \citenamefont {Bruin}, \citenamefont {Barber}, \citenamefont {Edkins},
  \citenamefont {Nishimura}, \citenamefont {Yonezawa}, \citenamefont {Maeno},\
  and\ \citenamefont {Mackenzie}}]{Hicks2014}%
  \BibitemOpen
  \bibfield  {author} {\bibinfo {author} {\bibfnamefont {C.~W.}\ \bibnamefont
  {Hicks}}, \bibinfo {author} {\bibfnamefont {D.~O.}\ \bibnamefont {Brodsky}},
  \bibinfo {author} {\bibfnamefont {E.~A.}\ \bibnamefont {Yelland}}, \bibinfo
  {author} {\bibfnamefont {A.~S.}\ \bibnamefont {Gibbs}}, \bibinfo {author}
  {\bibfnamefont {J.~A.~N.}\ \bibnamefont {Bruin}}, \bibinfo {author}
  {\bibfnamefont {M.~E.}\ \bibnamefont {Barber}}, \bibinfo {author}
  {\bibfnamefont {S.~D.}\ \bibnamefont {Edkins}}, \bibinfo {author}
  {\bibfnamefont {K.}~\bibnamefont {Nishimura}}, \bibinfo {author}
  {\bibfnamefont {S.}~\bibnamefont {Yonezawa}}, \bibinfo {author}
  {\bibfnamefont {Y.}~\bibnamefont {Maeno}}, \ and\ \bibinfo {author}
  {\bibfnamefont {A.~P.}\ \bibnamefont {Mackenzie}},\ }\href {\doibase
  10.1126/science.1248292} {\bibfield  {journal} {\bibinfo  {journal}
  {Science}\ }\textbf {\bibinfo {volume} {344}},\ \bibinfo {pages} {283}
  (\bibinfo {year} {2014})}\BibitemShut {NoStop}%
\bibitem [{\citenamefont {Steppke}\ \emph {et~al.}(2017)\citenamefont
  {Steppke}, \citenamefont {Zhao}, \citenamefont {Barber}, \citenamefont
  {Scaffidi}, \citenamefont {Jerzembeck}, \citenamefont {Rosner}, \citenamefont
  {Gibbs}, \citenamefont {Maeno}, \citenamefont {Simon}, \citenamefont
  {Mackenzie},\ and\ \citenamefont {Hicks}}]{Steppke2017}%
  \BibitemOpen
  \bibfield  {author} {\bibinfo {author} {\bibfnamefont {A.}~\bibnamefont
  {Steppke}}, \bibinfo {author} {\bibfnamefont {L.}~\bibnamefont {Zhao}},
  \bibinfo {author} {\bibfnamefont {M.~E.}\ \bibnamefont {Barber}}, \bibinfo
  {author} {\bibfnamefont {T.}~\bibnamefont {Scaffidi}}, \bibinfo {author}
  {\bibfnamefont {F.}~\bibnamefont {Jerzembeck}}, \bibinfo {author}
  {\bibfnamefont {H.}~\bibnamefont {Rosner}}, \bibinfo {author} {\bibfnamefont
  {A.~S.}\ \bibnamefont {Gibbs}}, \bibinfo {author} {\bibfnamefont
  {Y.}~\bibnamefont {Maeno}}, \bibinfo {author} {\bibfnamefont {S.~H.}\
  \bibnamefont {Simon}}, \bibinfo {author} {\bibfnamefont {A.~P.}\ \bibnamefont
  {Mackenzie}}, \ and\ \bibinfo {author} {\bibfnamefont {C.~W.}\ \bibnamefont
  {Hicks}},\ }\href {\doibase 10.1126/science.aaf9398} {\bibfield  {journal}
  {\bibinfo  {journal} {Science}\ }\textbf {\bibinfo {volume} {355}} (\bibinfo
  {year} {2017}),\ 10.1126/science.aaf9398}\BibitemShut {NoStop}%
\bibitem [{\citenamefont {Burganov}\ \emph {et~al.}(2016)\citenamefont
  {Burganov}, \citenamefont {Adamo}, \citenamefont {Mulder}, \citenamefont
  {Uchida}, \citenamefont {King}, \citenamefont {Harter}, \citenamefont {Shai},
  \citenamefont {Gibbs}, \citenamefont {Mackenzie}, \citenamefont {Uecker},
  \citenamefont {Bruetzam}, \citenamefont {Beasley}, \citenamefont {Fennie},
  \citenamefont {Schlom},\ and\ \citenamefont {Shen}}]{Burganov2016}%
  \BibitemOpen
  \bibfield  {author} {\bibinfo {author} {\bibfnamefont {B.}~\bibnamefont
  {Burganov}}, \bibinfo {author} {\bibfnamefont {C.}~\bibnamefont {Adamo}},
  \bibinfo {author} {\bibfnamefont {A.}~\bibnamefont {Mulder}}, \bibinfo
  {author} {\bibfnamefont {M.}~\bibnamefont {Uchida}}, \bibinfo {author}
  {\bibfnamefont {P.~D.~C.}\ \bibnamefont {King}}, \bibinfo {author}
  {\bibfnamefont {J.~W.}\ \bibnamefont {Harter}}, \bibinfo {author}
  {\bibfnamefont {D.~E.}\ \bibnamefont {Shai}}, \bibinfo {author}
  {\bibfnamefont {A.~S.}\ \bibnamefont {Gibbs}}, \bibinfo {author}
  {\bibfnamefont {A.~P.}\ \bibnamefont {Mackenzie}}, \bibinfo {author}
  {\bibfnamefont {R.}~\bibnamefont {Uecker}}, \bibinfo {author} {\bibfnamefont
  {M.}~\bibnamefont {Bruetzam}}, \bibinfo {author} {\bibfnamefont {M.~R.}\
  \bibnamefont {Beasley}}, \bibinfo {author} {\bibfnamefont {C.~J.}\
  \bibnamefont {Fennie}}, \bibinfo {author} {\bibfnamefont {D.~G.}\
  \bibnamefont {Schlom}}, \ and\ \bibinfo {author} {\bibfnamefont {K.~M.}\
  \bibnamefont {Shen}},\ }\href {\doibase 10.1103/PhysRevLett.116.197003}
  {\bibfield  {journal} {\bibinfo  {journal} {Phys. Rev. Lett.}\ }\textbf
  {\bibinfo {volume} {116}},\ \bibinfo {pages} {197003} (\bibinfo {year}
  {2016})}\BibitemShut {NoStop}%
\bibitem [{\citenamefont {Barber}\ \emph {et~al.}(2018)\citenamefont {Barber},
  \citenamefont {Gibbs}, \citenamefont {Maeno}, \citenamefont {Mackenzie},\
  and\ \citenamefont {Hicks}}]{Barber2018}%
  \BibitemOpen
  \bibfield  {author} {\bibinfo {author} {\bibfnamefont {M.~E.}\ \bibnamefont
  {Barber}}, \bibinfo {author} {\bibfnamefont {A.~S.}\ \bibnamefont {Gibbs}},
  \bibinfo {author} {\bibfnamefont {Y.}~\bibnamefont {Maeno}}, \bibinfo
  {author} {\bibfnamefont {A.~P.}\ \bibnamefont {Mackenzie}}, \ and\ \bibinfo
  {author} {\bibfnamefont {C.~W.}\ \bibnamefont {Hicks}},\ }\href {\doibase
  10.1103/PhysRevLett.120.076602} {\bibfield  {journal} {\bibinfo  {journal}
  {Phys. Rev. Lett.}\ }\textbf {\bibinfo {volume} {120}},\ \bibinfo {pages}
  {076602} (\bibinfo {year} {2018})}\BibitemShut {NoStop}%
\bibitem [{\citenamefont {Luo}\ \emph {et~al.}(2019)\citenamefont {Luo},
  \citenamefont {Pustogow}, \citenamefont {Guzman}, \citenamefont {Dioguardi},
  \citenamefont {Thomas}, \citenamefont {Ronning}, \citenamefont {Kikugawa},
  \citenamefont {Sokolov}, \citenamefont {Jerzembeck}, \citenamefont
  {Mackenzie}, \citenamefont {Hicks}, \citenamefont {Bauer}, \citenamefont
  {Mazin},\ and\ \citenamefont {Brown}}]{Luo2019}%
  \BibitemOpen
  \bibfield  {author} {\bibinfo {author} {\bibfnamefont {Y.}~\bibnamefont
  {Luo}}, \bibinfo {author} {\bibfnamefont {A.}~\bibnamefont {Pustogow}},
  \bibinfo {author} {\bibfnamefont {P.}~\bibnamefont {Guzman}}, \bibinfo
  {author} {\bibfnamefont {A.~P.}\ \bibnamefont {Dioguardi}}, \bibinfo {author}
  {\bibfnamefont {S.~M.}\ \bibnamefont {Thomas}}, \bibinfo {author}
  {\bibfnamefont {F.}~\bibnamefont {Ronning}}, \bibinfo {author} {\bibfnamefont
  {N.}~\bibnamefont {Kikugawa}}, \bibinfo {author} {\bibfnamefont {D.~A.}\
  \bibnamefont {Sokolov}}, \bibinfo {author} {\bibfnamefont {F.}~\bibnamefont
  {Jerzembeck}}, \bibinfo {author} {\bibfnamefont {A.~P.}\ \bibnamefont
  {Mackenzie}}, \bibinfo {author} {\bibfnamefont {C.~W.}\ \bibnamefont
  {Hicks}}, \bibinfo {author} {\bibfnamefont {E.~D.}\ \bibnamefont {Bauer}},
  \bibinfo {author} {\bibfnamefont {I.~I.}\ \bibnamefont {Mazin}}, \ and\
  \bibinfo {author} {\bibfnamefont {S.~E.}\ \bibnamefont {Brown}},\ }\href
  {\doibase 10.1103/PhysRevX.9.021044} {\bibfield  {journal} {\bibinfo
  {journal} {Phys. Rev. X}\ }\textbf {\bibinfo {volume} {9}},\ \bibinfo {pages}
  {021044} (\bibinfo {year} {2019})}\BibitemShut {NoStop}%
\bibitem [{\citenamefont {Steffens}\ \emph {et~al.}(2019)\citenamefont
  {Steffens}, \citenamefont {Sidis}, \citenamefont {Kulda}, \citenamefont
  {Mao}, \citenamefont {Maeno}, \citenamefont {Mazin},\ and\ \citenamefont
  {Braden}}]{Steffens2019}%
  \BibitemOpen
  \bibfield  {author} {\bibinfo {author} {\bibfnamefont {P.}~\bibnamefont
  {Steffens}}, \bibinfo {author} {\bibfnamefont {Y.}~\bibnamefont {Sidis}},
  \bibinfo {author} {\bibfnamefont {J.}~\bibnamefont {Kulda}}, \bibinfo
  {author} {\bibfnamefont {Z.~Q.}\ \bibnamefont {Mao}}, \bibinfo {author}
  {\bibfnamefont {Y.}~\bibnamefont {Maeno}}, \bibinfo {author} {\bibfnamefont
  {I.~I.}\ \bibnamefont {Mazin}}, \ and\ \bibinfo {author} {\bibfnamefont
  {M.}~\bibnamefont {Braden}},\ }\href {\doibase
  10.1103/PhysRevLett.122.047004} {\bibfield  {journal} {\bibinfo  {journal}
  {Phys. Rev. Lett.}\ }\textbf {\bibinfo {volume} {122}},\ \bibinfo {pages}
  {047004} (\bibinfo {year} {2019})}\BibitemShut {NoStop}%
\bibitem [{\citenamefont {Hsu}\ \emph {et~al.}(2016)\citenamefont {Hsu},
  \citenamefont {Cho}, \citenamefont {Rebola}, \citenamefont {Burganov},
  \citenamefont {Adamo}, \citenamefont {Shen}, \citenamefont {Schlom},
  \citenamefont {Fennie},\ and\ \citenamefont {Kim}}]{Hsu2016}%
  \BibitemOpen
  \bibfield  {author} {\bibinfo {author} {\bibfnamefont {Y.-T.}\ \bibnamefont
  {Hsu}}, \bibinfo {author} {\bibfnamefont {W.}~\bibnamefont {Cho}}, \bibinfo
  {author} {\bibfnamefont {A.~F.}\ \bibnamefont {Rebola}}, \bibinfo {author}
  {\bibfnamefont {B.}~\bibnamefont {Burganov}}, \bibinfo {author}
  {\bibfnamefont {C.}~\bibnamefont {Adamo}}, \bibinfo {author} {\bibfnamefont
  {K.~M.}\ \bibnamefont {Shen}}, \bibinfo {author} {\bibfnamefont {D.~G.}\
  \bibnamefont {Schlom}}, \bibinfo {author} {\bibfnamefont {C.~J.}\
  \bibnamefont {Fennie}}, \ and\ \bibinfo {author} {\bibfnamefont {E.-A.}\
  \bibnamefont {Kim}},\ }\href {\doibase 10.1103/PhysRevB.94.045118} {\bibfield
   {journal} {\bibinfo  {journal} {Phys. Rev. B}\ }\textbf {\bibinfo {volume}
  {94}},\ \bibinfo {pages} {045118} (\bibinfo {year} {2016})}\BibitemShut
  {NoStop}%
\bibitem [{\citenamefont {Liu}\ \emph {et~al.}(2018)\citenamefont {Liu},
  \citenamefont {Wang}, \citenamefont {Zhang},\ and\ \citenamefont
  {Wang}}]{Liu2018}%
  \BibitemOpen
  \bibfield  {author} {\bibinfo {author} {\bibfnamefont {Y.-C.}\ \bibnamefont
  {Liu}}, \bibinfo {author} {\bibfnamefont {W.-S.}\ \bibnamefont {Wang}},
  \bibinfo {author} {\bibfnamefont {F.-C.}\ \bibnamefont {Zhang}}, \ and\
  \bibinfo {author} {\bibfnamefont {Q.-H.}\ \bibnamefont {Wang}},\ }\href
  {\doibase 10.1103/PhysRevB.97.224522} {\bibfield  {journal} {\bibinfo
  {journal} {Phys. Rev. B}\ }\textbf {\bibinfo {volume} {97}},\ \bibinfo
  {pages} {224522} (\bibinfo {year} {2018})}\BibitemShut {NoStop}%
\bibitem [{\citenamefont {Arita}\ \emph {et~al.}(2007)\citenamefont {Arita},
  \citenamefont {Yamasaki}, \citenamefont {Held}, \citenamefont {Matsuno},\
  and\ \citenamefont {Kuroki}}]{Arita2007_1}%
  \BibitemOpen
  \bibfield  {author} {\bibinfo {author} {\bibfnamefont {R.}~\bibnamefont
  {Arita}}, \bibinfo {author} {\bibfnamefont {A.}~\bibnamefont {Yamasaki}},
  \bibinfo {author} {\bibfnamefont {K.}~\bibnamefont {Held}}, \bibinfo {author}
  {\bibfnamefont {J.}~\bibnamefont {Matsuno}}, \ and\ \bibinfo {author}
  {\bibfnamefont {K.}~\bibnamefont {Kuroki}},\ }\href {\doibase
  10.1103/PhysRevB.75.174521} {\bibfield  {journal} {\bibinfo  {journal} {Phys.
  Rev. B}\ }\textbf {\bibinfo {volume} {75}},\ \bibinfo {pages} {174521}
  (\bibinfo {year} {2007})}\BibitemShut {NoStop}%
\bibitem [{\citenamefont {Anisimov}\ \emph {et~al.}(1999)\citenamefont
  {Anisimov}, \citenamefont {Bukhvalov},\ and\ \citenamefont
  {Rice}}]{Anisimov1999}%
  \BibitemOpen
  \bibfield  {author} {\bibinfo {author} {\bibfnamefont {V.~I.}\ \bibnamefont
  {Anisimov}}, \bibinfo {author} {\bibfnamefont {D.}~\bibnamefont {Bukhvalov}},
  \ and\ \bibinfo {author} {\bibfnamefont {T.~M.}\ \bibnamefont {Rice}},\
  }\href {\doibase 10.1103/PhysRevB.59.7901} {\bibfield  {journal} {\bibinfo
  {journal} {Phys. Rev. B}\ }\textbf {\bibinfo {volume} {59}},\ \bibinfo
  {pages} {7901} (\bibinfo {year} {1999})}\BibitemShut {NoStop}%
\bibitem [{\citenamefont {Chaloupka}\ and\ \citenamefont
  {Khaliullin}(2008)}]{Chaloupka2008}%
  \BibitemOpen
  \bibfield  {author} {\bibinfo {author} {\bibfnamefont {J.}~\bibnamefont
  {Chaloupka}}\ and\ \bibinfo {author} {\bibfnamefont {G.}~\bibnamefont
  {Khaliullin}},\ }\href {\doibase 10.1103/PhysRevLett.100.016404} {\bibfield
  {journal} {\bibinfo  {journal} {Phys. Rev. Lett.}\ }\textbf {\bibinfo
  {volume} {100}},\ \bibinfo {pages} {016404} (\bibinfo {year}
  {2008})}\BibitemShut {NoStop}%
\bibitem [{\citenamefont {Hansmann}\ \emph {et~al.}(2009)\citenamefont
  {Hansmann}, \citenamefont {Yang}, \citenamefont {Toschi}, \citenamefont
  {Khaliullin}, \citenamefont {Andersen},\ and\ \citenamefont
  {Held}}]{Hansmann2009}%
  \BibitemOpen
  \bibfield  {author} {\bibinfo {author} {\bibfnamefont {P.}~\bibnamefont
  {Hansmann}}, \bibinfo {author} {\bibfnamefont {X.}~\bibnamefont {Yang}},
  \bibinfo {author} {\bibfnamefont {A.}~\bibnamefont {Toschi}}, \bibinfo
  {author} {\bibfnamefont {G.}~\bibnamefont {Khaliullin}}, \bibinfo {author}
  {\bibfnamefont {O.~K.}\ \bibnamefont {Andersen}}, \ and\ \bibinfo {author}
  {\bibfnamefont {K.}~\bibnamefont {Held}},\ }\href {\doibase
  10.1103/PhysRevLett.103.016401} {\bibfield  {journal} {\bibinfo  {journal}
  {Phys. Rev. Lett.}\ }\textbf {\bibinfo {volume} {103}},\ \bibinfo {pages}
  {016401} (\bibinfo {year} {2009})}\BibitemShut {NoStop}%
\bibitem [{\citenamefont {Watanabe}\ \emph {et~al.}(2013)\citenamefont
  {Watanabe}, \citenamefont {Shirakawa},\ and\ \citenamefont
  {Yunoki}}]{Watanabe2013}%
  \BibitemOpen
  \bibfield  {author} {\bibinfo {author} {\bibfnamefont {H.}~\bibnamefont
  {Watanabe}}, \bibinfo {author} {\bibfnamefont {T.}~\bibnamefont {Shirakawa}},
  \ and\ \bibinfo {author} {\bibfnamefont {S.}~\bibnamefont {Yunoki}},\ }\href
  {\doibase 10.1103/PhysRevLett.110.027002} {\bibfield  {journal} {\bibinfo
  {journal} {Phys. Rev. Lett.}\ }\textbf {\bibinfo {volume} {110}},\ \bibinfo
  {pages} {027002} (\bibinfo {year} {2013})}\BibitemShut {NoStop}%
\bibitem [{\citenamefont {Kim}\ \emph {et~al.}(2014)\citenamefont {Kim},
  \citenamefont {Krupin}, , \citenamefont {Denlinger}, \citenamefont
  {Bostwick}, \citenamefont {Rotenberg}, \citenamefont {Zhao}, \citenamefont
  {Mitchell}, \citenamefont {Allen},\ and\ \citenamefont {Kim}}]{YKKim2014}%
  \BibitemOpen
  \bibfield  {author} {\bibinfo {author} {\bibfnamefont {Y.~K.}\ \bibnamefont
  {Kim}}, \bibinfo {author} {\bibfnamefont {O.}~\bibnamefont {Krupin}}, ,
  \bibinfo {author} {\bibfnamefont {J.~D.}\ \bibnamefont {Denlinger}}, \bibinfo
  {author} {\bibfnamefont {A.}~\bibnamefont {Bostwick}}, \bibinfo {author}
  {\bibfnamefont {E.}~\bibnamefont {Rotenberg}}, \bibinfo {author}
  {\bibfnamefont {Q.}~\bibnamefont {Zhao}}, \bibinfo {author} {\bibfnamefont
  {J.~F.}\ \bibnamefont {Mitchell}}, \bibinfo {author} {\bibfnamefont {J.~W.}\
  \bibnamefont {Allen}}, \ and\ \bibinfo {author} {\bibfnamefont {B.~J.}\
  \bibnamefont {Kim}},\ }\href@noop {} {\bibfield  {journal} {\bibinfo
  {journal} {Science}\ }\textbf {\bibinfo {volume} {345}},\ \bibinfo {pages}
  {187} (\bibinfo {year} {2014})}\BibitemShut {NoStop}%
\bibitem [{\citenamefont {Gawraczy{\'n}ski}\ \emph {et~al.}(2019)\citenamefont
  {Gawraczy{\'n}ski}, \citenamefont {Kurzyd{\l}owski}, \citenamefont {Ewings},
  \citenamefont {Bandaru}, \citenamefont {Gadomski}, \citenamefont {Mazej},
  \citenamefont {Ruani}, \citenamefont {Bergenti}, \citenamefont {Jaro{\'n}},
  \citenamefont {Ozarowski}, \citenamefont {Hill}, \citenamefont
  {Leszczy{\'n}ski}, \citenamefont {Tok{\'a}r}, \citenamefont {Derzsi},
  \citenamefont {Barone}, \citenamefont {Wohlfeld}, \citenamefont {Lorenzana},\
  and\ \citenamefont {Grochala}}]{Gaw2019}%
  \BibitemOpen
  \bibfield  {author} {\bibinfo {author} {\bibfnamefont {J.}~\bibnamefont
  {Gawraczy{\'n}ski}}, \bibinfo {author} {\bibfnamefont {D.}~\bibnamefont
  {Kurzyd{\l}owski}}, \bibinfo {author} {\bibfnamefont {R.~A.}\ \bibnamefont
  {Ewings}}, \bibinfo {author} {\bibfnamefont {S.}~\bibnamefont {Bandaru}},
  \bibinfo {author} {\bibfnamefont {W.}~\bibnamefont {Gadomski}}, \bibinfo
  {author} {\bibfnamefont {Z.}~\bibnamefont {Mazej}}, \bibinfo {author}
  {\bibfnamefont {G.}~\bibnamefont {Ruani}}, \bibinfo {author} {\bibfnamefont
  {I.}~\bibnamefont {Bergenti}}, \bibinfo {author} {\bibfnamefont
  {T.}~\bibnamefont {Jaro{\'n}}}, \bibinfo {author} {\bibfnamefont
  {A.}~\bibnamefont {Ozarowski}}, \bibinfo {author} {\bibfnamefont
  {S.}~\bibnamefont {Hill}}, \bibinfo {author} {\bibfnamefont {P.~J.}\
  \bibnamefont {Leszczy{\'n}ski}}, \bibinfo {author} {\bibfnamefont
  {K.}~\bibnamefont {Tok{\'a}r}}, \bibinfo {author} {\bibfnamefont
  {M.}~\bibnamefont {Derzsi}}, \bibinfo {author} {\bibfnamefont
  {P.}~\bibnamefont {Barone}}, \bibinfo {author} {\bibfnamefont
  {K.}~\bibnamefont {Wohlfeld}}, \bibinfo {author} {\bibfnamefont
  {J.}~\bibnamefont {Lorenzana}}, \ and\ \bibinfo {author} {\bibfnamefont
  {W.}~\bibnamefont {Grochala}},\ }\href {\doibase 10.1073/pnas.1812857116}
  {\bibfield  {journal} {\bibinfo  {journal} {Proceedings of the National
  Academy of Sciences}\ }\textbf {\bibinfo {volume} {116}},\ \bibinfo {pages}
  {1495} (\bibinfo {year} {2019})}\BibitemShut {NoStop}%
\bibitem [{\citenamefont {Isaacs}\ and\ \citenamefont
  {Wolverton}(2019)}]{Isaacs2019}%
  \BibitemOpen
  \bibfield  {author} {\bibinfo {author} {\bibfnamefont {E.~B.}\ \bibnamefont
  {Isaacs}}\ and\ \bibinfo {author} {\bibfnamefont {C.}~\bibnamefont
  {Wolverton}},\ }\href {\doibase 10.1103/PhysRevX.9.021042} {\bibfield
  {journal} {\bibinfo  {journal} {Phys. Rev. X}\ }\textbf {\bibinfo {volume}
  {9}},\ \bibinfo {pages} {021042} (\bibinfo {year} {2019})}\BibitemShut
  {NoStop}%
\bibitem [{\citenamefont {Li}\ \emph {et~al.}(2019)\citenamefont {Li},
  \citenamefont {Lee}, \citenamefont {Wang}, \citenamefont {Osada},
  \citenamefont {Crossley}, \citenamefont {Lee}, \citenamefont {Cui},
  \citenamefont {Hikita},\ and\ \citenamefont {Hwang}}]{Li2019}%
  \BibitemOpen
  \bibfield  {author} {\bibinfo {author} {\bibfnamefont {D.}~\bibnamefont
  {Li}}, \bibinfo {author} {\bibfnamefont {K.}~\bibnamefont {Lee}}, \bibinfo
  {author} {\bibfnamefont {B.~Y.}\ \bibnamefont {Wang}}, \bibinfo {author}
  {\bibfnamefont {M.}~\bibnamefont {Osada}}, \bibinfo {author} {\bibfnamefont
  {S.}~\bibnamefont {Crossley}}, \bibinfo {author} {\bibfnamefont {H.~R.}\
  \bibnamefont {Lee}}, \bibinfo {author} {\bibfnamefont {Y.}~\bibnamefont
  {Cui}}, \bibinfo {author} {\bibfnamefont {Y.}~\bibnamefont {Hikita}}, \ and\
  \bibinfo {author} {\bibfnamefont {H.~Y.}\ \bibnamefont {Hwang}},\ }\href@noop
  {} {\bibfield  {journal} {\bibinfo  {journal} {Nature}\ }\textbf {\bibinfo
  {volume} {572}},\ \bibinfo {pages} {624} (\bibinfo {year}
  {2019})}\BibitemShut {NoStop}%
\bibitem [{\citenamefont {Koster}\ \emph {et~al.}(2012)\citenamefont {Koster},
  \citenamefont {Klein}, \citenamefont {Siemons}, \citenamefont {Rijnders},
  \citenamefont {Dodge}, \citenamefont {Eom}, \citenamefont {Blank},\ and\
  \citenamefont {Beasley}}]{Koster2012}%
  \BibitemOpen
  \bibfield  {author} {\bibinfo {author} {\bibfnamefont {G.}~\bibnamefont
  {Koster}}, \bibinfo {author} {\bibfnamefont {L.}~\bibnamefont {Klein}},
  \bibinfo {author} {\bibfnamefont {W.}~\bibnamefont {Siemons}}, \bibinfo
  {author} {\bibfnamefont {G.}~\bibnamefont {Rijnders}}, \bibinfo {author}
  {\bibfnamefont {J.~S.}\ \bibnamefont {Dodge}}, \bibinfo {author}
  {\bibfnamefont {C.-B.}\ \bibnamefont {Eom}}, \bibinfo {author} {\bibfnamefont
  {D.~H.~A.}\ \bibnamefont {Blank}}, \ and\ \bibinfo {author} {\bibfnamefont
  {M.~R.}\ \bibnamefont {Beasley}},\ }\href {\doibase
  10.1103/RevModPhys.84.253} {\bibfield  {journal} {\bibinfo  {journal} {Rev.
  Mod. Phys.}\ }\textbf {\bibinfo {volume} {84}},\ \bibinfo {pages} {253}
  (\bibinfo {year} {2012})}\BibitemShut {NoStop}%
\bibitem [{\citenamefont {Gu}\ \emph {et~al.}(2012)\citenamefont {Gu},
  \citenamefont {Xie}, \citenamefont {Shen}, \citenamefont {Xie}, \citenamefont
  {Wang}, \citenamefont {Tang}, \citenamefont {Wu}, \citenamefont {Zhang},\
  and\ \citenamefont {Wu}}]{Gu2012}%
  \BibitemOpen
  \bibfield  {author} {\bibinfo {author} {\bibfnamefont {M.}~\bibnamefont
  {Gu}}, \bibinfo {author} {\bibfnamefont {Q.}~\bibnamefont {Xie}}, \bibinfo
  {author} {\bibfnamefont {X.}~\bibnamefont {Shen}}, \bibinfo {author}
  {\bibfnamefont {R.}~\bibnamefont {Xie}}, \bibinfo {author} {\bibfnamefont
  {J.}~\bibnamefont {Wang}}, \bibinfo {author} {\bibfnamefont {G.}~\bibnamefont
  {Tang}}, \bibinfo {author} {\bibfnamefont {D.}~\bibnamefont {Wu}}, \bibinfo
  {author} {\bibfnamefont {G.~P.}\ \bibnamefont {Zhang}}, \ and\ \bibinfo
  {author} {\bibfnamefont {X.~S.}\ \bibnamefont {Wu}},\ }\href {\doibase
  10.1103/PhysRevLett.109.157003} {\bibfield  {journal} {\bibinfo  {journal}
  {Phys. Rev. Lett.}\ }\textbf {\bibinfo {volume} {109}},\ \bibinfo {pages}
  {157003} (\bibinfo {year} {2012})}\BibitemShut {NoStop}%
\bibitem [{\citenamefont {Boschker}\ \emph {et~al.}(2019)\citenamefont
  {Boschker}, \citenamefont {Harada}, \citenamefont {Asaba}, \citenamefont
  {Ashoori}, \citenamefont {Boris}, \citenamefont {Hilgenkamp}, \citenamefont
  {Hughes}, \citenamefont {Holtz}, \citenamefont {Li}, \citenamefont {Muller},
  \citenamefont {Nair}, \citenamefont {Reith}, \citenamefont {Renshaw~Wang},
  \citenamefont {Schlom}, \citenamefont {Soukiassian},\ and\ \citenamefont
  {Mannhart}}]{Boschker2019}%
  \BibitemOpen
  \bibfield  {author} {\bibinfo {author} {\bibfnamefont {H.}~\bibnamefont
  {Boschker}}, \bibinfo {author} {\bibfnamefont {T.}~\bibnamefont {Harada}},
  \bibinfo {author} {\bibfnamefont {T.}~\bibnamefont {Asaba}}, \bibinfo
  {author} {\bibfnamefont {R.}~\bibnamefont {Ashoori}}, \bibinfo {author}
  {\bibfnamefont {A.~V.}\ \bibnamefont {Boris}}, \bibinfo {author}
  {\bibfnamefont {H.}~\bibnamefont {Hilgenkamp}}, \bibinfo {author}
  {\bibfnamefont {C.~R.}\ \bibnamefont {Hughes}}, \bibinfo {author}
  {\bibfnamefont {M.~E.}\ \bibnamefont {Holtz}}, \bibinfo {author}
  {\bibfnamefont {L.}~\bibnamefont {Li}}, \bibinfo {author} {\bibfnamefont
  {D.~A.}\ \bibnamefont {Muller}}, \bibinfo {author} {\bibfnamefont
  {H.}~\bibnamefont {Nair}}, \bibinfo {author} {\bibfnamefont {P.}~\bibnamefont
  {Reith}}, \bibinfo {author} {\bibfnamefont {X.}~\bibnamefont {Renshaw~Wang}},
  \bibinfo {author} {\bibfnamefont {D.~G.}\ \bibnamefont {Schlom}}, \bibinfo
  {author} {\bibfnamefont {A.}~\bibnamefont {Soukiassian}}, \ and\ \bibinfo
  {author} {\bibfnamefont {J.}~\bibnamefont {Mannhart}},\ }\href {\doibase
  10.1103/PhysRevX.9.011027} {\bibfield  {journal} {\bibinfo  {journal} {Phys.
  Rev. X}\ }\textbf {\bibinfo {volume} {9}},\ \bibinfo {pages} {011027}
  (\bibinfo {year} {2019})}\BibitemShut {NoStop}%
\bibitem [{\citenamefont {Jeong}\ \emph {et~al.}(2020)\citenamefont {Jeong},
  \citenamefont {Min}, \citenamefont {Woo}, \citenamefont {Kim}, \citenamefont
  {Zhang}, \citenamefont {Cho}, \citenamefont {Son}, \citenamefont {Kim},
  \citenamefont {Han}, \citenamefont {Park}, \citenamefont {Jeong},
  \citenamefont {Ohta}, \citenamefont {Lee}, \citenamefont {Noh}, \citenamefont
  {Lee},\ and\ \citenamefont {Choi}}]{Jeong2020}%
  \BibitemOpen
  \bibfield  {author} {\bibinfo {author} {\bibfnamefont {S.~G.}\ \bibnamefont
  {Jeong}}, \bibinfo {author} {\bibfnamefont {T.}~\bibnamefont {Min}}, \bibinfo
  {author} {\bibfnamefont {S.}~\bibnamefont {Woo}}, \bibinfo {author}
  {\bibfnamefont {J.}~\bibnamefont {Kim}}, \bibinfo {author} {\bibfnamefont
  {Y.-Q.}\ \bibnamefont {Zhang}}, \bibinfo {author} {\bibfnamefont {S.~W.}\
  \bibnamefont {Cho}}, \bibinfo {author} {\bibfnamefont {J.}~\bibnamefont
  {Son}}, \bibinfo {author} {\bibfnamefont {Y.-M.}\ \bibnamefont {Kim}},
  \bibinfo {author} {\bibfnamefont {J.~H.}\ \bibnamefont {Han}}, \bibinfo
  {author} {\bibfnamefont {S.}~\bibnamefont {Park}}, \bibinfo {author}
  {\bibfnamefont {H.~Y.}\ \bibnamefont {Jeong}}, \bibinfo {author}
  {\bibfnamefont {H.}~\bibnamefont {Ohta}}, \bibinfo {author} {\bibfnamefont
  {S.}~\bibnamefont {Lee}}, \bibinfo {author} {\bibfnamefont {T.~W.}\
  \bibnamefont {Noh}}, \bibinfo {author} {\bibfnamefont {J.}~\bibnamefont
  {Lee}}, \ and\ \bibinfo {author} {\bibfnamefont {W.~S.}\ \bibnamefont
  {Choi}},\ }\href {\doibase 10.1103/PhysRevLett.124.026401} {\bibfield
  {journal} {\bibinfo  {journal} {Phys. Rev. Lett.}\ }\textbf {\bibinfo
  {volume} {124}},\ \bibinfo {pages} {026401} (\bibinfo {year}
  {2020})}\BibitemShut {NoStop}%
\bibitem [{\citenamefont {Matzdorf}\ \emph {et~al.}(2000)\citenamefont
  {Matzdorf}, \citenamefont {Fang}, \citenamefont {Ismail}, \citenamefont
  {Zhang}, \citenamefont {Kimura}, \citenamefont {Y.~Tokura~and},\ and\
  \citenamefont {Plummer}}]{Matzdorf2000}%
  \BibitemOpen
  \bibfield  {author} {\bibinfo {author} {\bibfnamefont {R.}~\bibnamefont
  {Matzdorf}}, \bibinfo {author} {\bibfnamefont {Z.}~\bibnamefont {Fang}},
  \bibinfo {author} {\bibnamefont {Ismail}}, \bibinfo {author} {\bibfnamefont
  {J.}~\bibnamefont {Zhang}}, \bibinfo {author} {\bibfnamefont
  {T.}~\bibnamefont {Kimura}}, \bibinfo {author} {\bibfnamefont {K.~T.}\
  \bibnamefont {Y.~Tokura~and}}, \ and\ \bibinfo {author} {\bibfnamefont
  {E.~W.}\ \bibnamefont {Plummer}},\ }\href@noop {} {\bibfield  {journal}
  {\bibinfo  {journal} {Science}\ }\textbf {\bibinfo {volume} {289}},\ \bibinfo
  {pages} {746} (\bibinfo {year} {2000})}\BibitemShut {NoStop}%
\bibitem [{\citenamefont {Damascelli}\ \emph {et~al.}(2000)\citenamefont
  {Damascelli}, \citenamefont {Lu}, \citenamefont {Shen}, \citenamefont
  {Armitage}, \citenamefont {Ronning}, \citenamefont {Feng}, \citenamefont
  {Kim}, \citenamefont {Shen}, \citenamefont {Kimura}, \citenamefont {Tokura},
  \citenamefont {Mao},\ and\ \citenamefont {Maeno}}]{Damascelli2000}%
  \BibitemOpen
  \bibfield  {author} {\bibinfo {author} {\bibfnamefont {A.}~\bibnamefont
  {Damascelli}}, \bibinfo {author} {\bibfnamefont {D.~H.}\ \bibnamefont {Lu}},
  \bibinfo {author} {\bibfnamefont {K.~M.}\ \bibnamefont {Shen}}, \bibinfo
  {author} {\bibfnamefont {N.~P.}\ \bibnamefont {Armitage}}, \bibinfo {author}
  {\bibfnamefont {F.}~\bibnamefont {Ronning}}, \bibinfo {author} {\bibfnamefont
  {D.~L.}\ \bibnamefont {Feng}}, \bibinfo {author} {\bibfnamefont
  {C.}~\bibnamefont {Kim}}, \bibinfo {author} {\bibfnamefont {Z.-X.}\
  \bibnamefont {Shen}}, \bibinfo {author} {\bibfnamefont {T.}~\bibnamefont
  {Kimura}}, \bibinfo {author} {\bibfnamefont {Y.}~\bibnamefont {Tokura}},
  \bibinfo {author} {\bibfnamefont {Z.~Q.}\ \bibnamefont {Mao}}, \ and\
  \bibinfo {author} {\bibfnamefont {Y.}~\bibnamefont {Maeno}},\ }\href
  {\doibase 10.1103/PhysRevLett.85.5194} {\bibfield  {journal} {\bibinfo
  {journal} {Phys. Rev. Lett.}\ }\textbf {\bibinfo {volume} {85}},\ \bibinfo
  {pages} {5194} (\bibinfo {year} {2000})}\BibitemShut {NoStop}%
\bibitem [{\citenamefont {Wang}\ \emph {et~al.}(2004)\citenamefont {Wang},
  \citenamefont {Yang}, \citenamefont {Sekharan}, \citenamefont {Souma},
  \citenamefont {Matsui}, \citenamefont {Sato}, \citenamefont {Takahashi},
  \citenamefont {Lu}, \citenamefont {Zhang}, \citenamefont {Jin}, \citenamefont
  {Mandrus}, \citenamefont {Plummer}, \citenamefont {Wang},\ and\ \citenamefont
  {Ding}}]{Wang2004}%
  \BibitemOpen
  \bibfield  {author} {\bibinfo {author} {\bibfnamefont {S.-C.}\ \bibnamefont
  {Wang}}, \bibinfo {author} {\bibfnamefont {H.-B.}\ \bibnamefont {Yang}},
  \bibinfo {author} {\bibfnamefont {A.~K.~P.}\ \bibnamefont {Sekharan}},
  \bibinfo {author} {\bibfnamefont {S.}~\bibnamefont {Souma}}, \bibinfo
  {author} {\bibfnamefont {H.}~\bibnamefont {Matsui}}, \bibinfo {author}
  {\bibfnamefont {T.}~\bibnamefont {Sato}}, \bibinfo {author} {\bibfnamefont
  {T.}~\bibnamefont {Takahashi}}, \bibinfo {author} {\bibfnamefont
  {C.}~\bibnamefont {Lu}}, \bibinfo {author} {\bibfnamefont {J.}~\bibnamefont
  {Zhang}}, \bibinfo {author} {\bibfnamefont {R.}~\bibnamefont {Jin}}, \bibinfo
  {author} {\bibfnamefont {D.}~\bibnamefont {Mandrus}}, \bibinfo {author}
  {\bibfnamefont {E.~W.}\ \bibnamefont {Plummer}}, \bibinfo {author}
  {\bibfnamefont {Z.}~\bibnamefont {Wang}}, \ and\ \bibinfo {author}
  {\bibfnamefont {H.}~\bibnamefont {Ding}},\ }\href {\doibase
  10.1103/PhysRevLett.93.177007} {\bibfield  {journal} {\bibinfo  {journal}
  {Phys. Rev. Lett.}\ }\textbf {\bibinfo {volume} {93}},\ \bibinfo {pages}
  {177007} (\bibinfo {year} {2004})}\BibitemShut {NoStop}%
\bibitem [{\citenamefont {Firmo}\ \emph {et~al.}(2013)\citenamefont {Firmo},
  \citenamefont {Lederer}, \citenamefont {Lupien}, \citenamefont {Mackenzie},
  \citenamefont {Davis},\ and\ \citenamefont {Kivelson}}]{Firmo2013}%
  \BibitemOpen
  \bibfield  {author} {\bibinfo {author} {\bibfnamefont {I.~A.}\ \bibnamefont
  {Firmo}}, \bibinfo {author} {\bibfnamefont {S.}~\bibnamefont {Lederer}},
  \bibinfo {author} {\bibfnamefont {C.}~\bibnamefont {Lupien}}, \bibinfo
  {author} {\bibfnamefont {A.~P.}\ \bibnamefont {Mackenzie}}, \bibinfo {author}
  {\bibfnamefont {J.~C.}\ \bibnamefont {Davis}}, \ and\ \bibinfo {author}
  {\bibfnamefont {S.~A.}\ \bibnamefont {Kivelson}},\ }\href {\doibase
  10.1103/PhysRevB.88.134521} {\bibfield  {journal} {\bibinfo  {journal} {Phys.
  Rev. B}\ }\textbf {\bibinfo {volume} {88}},\ \bibinfo {pages} {134521}
  (\bibinfo {year} {2013})}\BibitemShut {NoStop}%
\bibitem [{\citenamefont {Agterberg}\ \emph {et~al.}(1997)\citenamefont
  {Agterberg}, \citenamefont {Rice},\ and\ \citenamefont
  {Sigrist}}]{Agterberg1997}%
  \BibitemOpen
  \bibfield  {author} {\bibinfo {author} {\bibfnamefont {D.~F.}\ \bibnamefont
  {Agterberg}}, \bibinfo {author} {\bibfnamefont {T.~M.}\ \bibnamefont {Rice}},
  \ and\ \bibinfo {author} {\bibfnamefont {M.}~\bibnamefont {Sigrist}},\ }\href
  {\doibase 10.1103/PhysRevLett.78.3374} {\bibfield  {journal} {\bibinfo
  {journal} {Phys. Rev. Lett.}\ }\textbf {\bibinfo {volume} {78}},\ \bibinfo
  {pages} {3374} (\bibinfo {year} {1997})}\BibitemShut {NoStop}%
\bibitem [{\citenamefont {Yanase}\ \emph {et~al.}(2003)\citenamefont {Yanase},
  \citenamefont {Jujo}, \citenamefont {Nomura}, \citenamefont {Ikeda},
  \citenamefont {Hotta},\ and\ \citenamefont {Yamada}}]{Yanase2003}%
  \BibitemOpen
  \bibfield  {author} {\bibinfo {author} {\bibfnamefont {Y.}~\bibnamefont
  {Yanase}}, \bibinfo {author} {\bibfnamefont {T.}~\bibnamefont {Jujo}},
  \bibinfo {author} {\bibfnamefont {T.}~\bibnamefont {Nomura}}, \bibinfo
  {author} {\bibfnamefont {H.}~\bibnamefont {Ikeda}}, \bibinfo {author}
  {\bibfnamefont {T.}~\bibnamefont {Hotta}}, \ and\ \bibinfo {author}
  {\bibfnamefont {K.}~\bibnamefont {Yamada}},\ }\href {\doibase
  https://doi.org/10.1016/j.physrep.2003.07.002} {\bibfield  {journal}
  {\bibinfo  {journal} {Physics Reports}\ }\textbf {\bibinfo {volume} {387}},\
  \bibinfo {pages} {1 } (\bibinfo {year} {2003})}\BibitemShut {NoStop}%
\bibitem [{\citenamefont {Raghu}\ \emph {et~al.}(2010)\citenamefont {Raghu},
  \citenamefont {Kapitulnik},\ and\ \citenamefont {Kivelson}}]{Raghu2010}%
  \BibitemOpen
  \bibfield  {author} {\bibinfo {author} {\bibfnamefont {S.}~\bibnamefont
  {Raghu}}, \bibinfo {author} {\bibfnamefont {A.}~\bibnamefont {Kapitulnik}}, \
  and\ \bibinfo {author} {\bibfnamefont {S.~A.}\ \bibnamefont {Kivelson}},\
  }\href {\doibase 10.1103/PhysRevLett.105.136401} {\bibfield  {journal}
  {\bibinfo  {journal} {Phys. Rev. Lett.}\ }\textbf {\bibinfo {volume} {105}},\
  \bibinfo {pages} {136401} (\bibinfo {year} {2010})}\BibitemShut {NoStop}%
\bibitem [{\citenamefont {Mazin}\ and\ \citenamefont
  {Singh}(1997)}]{Mazin1997}%
  \BibitemOpen
  \bibfield  {author} {\bibinfo {author} {\bibfnamefont {I.~I.}\ \bibnamefont
  {Mazin}}\ and\ \bibinfo {author} {\bibfnamefont {D.~J.}\ \bibnamefont
  {Singh}},\ }\href {\doibase 10.1103/PhysRevLett.79.733} {\bibfield  {journal}
  {\bibinfo  {journal} {Phys. Rev. Lett.}\ }\textbf {\bibinfo {volume} {79}},\
  \bibinfo {pages} {733} (\bibinfo {year} {1997})}\BibitemShut {NoStop}%
\bibitem [{\citenamefont {Mazin}\ and\ \citenamefont
  {Singh}(1999)}]{Mazin1999}%
  \BibitemOpen
  \bibfield  {author} {\bibinfo {author} {\bibfnamefont {I.~I.}\ \bibnamefont
  {Mazin}}\ and\ \bibinfo {author} {\bibfnamefont {D.~J.}\ \bibnamefont
  {Singh}},\ }\href {\doibase 10.1103/PhysRevLett.82.4324} {\bibfield
  {journal} {\bibinfo  {journal} {Phys. Rev. Lett.}\ }\textbf {\bibinfo
  {volume} {82}},\ \bibinfo {pages} {4324} (\bibinfo {year}
  {1999})}\BibitemShut {NoStop}%
\bibitem [{\citenamefont {Mahadevan}\ \emph {et~al.}(2009)\citenamefont
  {Mahadevan}, \citenamefont {Aryasetiawan}, \citenamefont {Janotti},\ and\
  \citenamefont {Sasaki}}]{Mahadevan2009}%
  \BibitemOpen
  \bibfield  {author} {\bibinfo {author} {\bibfnamefont {P.}~\bibnamefont
  {Mahadevan}}, \bibinfo {author} {\bibfnamefont {F.}~\bibnamefont
  {Aryasetiawan}}, \bibinfo {author} {\bibfnamefont {A.}~\bibnamefont
  {Janotti}}, \ and\ \bibinfo {author} {\bibfnamefont {T.}~\bibnamefont
  {Sasaki}},\ }\href {\doibase 10.1103/PhysRevB.80.035106} {\bibfield
  {journal} {\bibinfo  {journal} {Phys. Rev. B}\ }\textbf {\bibinfo {volume}
  {80}},\ \bibinfo {pages} {035106} (\bibinfo {year} {2009})}\BibitemShut
  {NoStop}%
\bibitem [{\citenamefont {Verissimo-Alves}\ \emph {et~al.}(2012)\citenamefont
  {Verissimo-Alves}, \citenamefont {Garc\'{\i}a-Fern\'andez}, \citenamefont
  {Bilc}, \citenamefont {Ghosez},\ and\ \citenamefont {Junquera}}]{vAlves2012}%
  \BibitemOpen
  \bibfield  {author} {\bibinfo {author} {\bibfnamefont {M.}~\bibnamefont
  {Verissimo-Alves}}, \bibinfo {author} {\bibfnamefont {P.}~\bibnamefont
  {Garc\'{\i}a-Fern\'andez}}, \bibinfo {author} {\bibfnamefont {D.~I.}\
  \bibnamefont {Bilc}}, \bibinfo {author} {\bibfnamefont {P.}~\bibnamefont
  {Ghosez}}, \ and\ \bibinfo {author} {\bibfnamefont {J.}~\bibnamefont
  {Junquera}},\ }\href {\doibase 10.1103/PhysRevLett.108.107003} {\bibfield
  {journal} {\bibinfo  {journal} {Phys. Rev. Lett.}\ }\textbf {\bibinfo
  {volume} {108}},\ \bibinfo {pages} {107003} (\bibinfo {year}
  {2012})}\BibitemShut {NoStop}%
\bibitem [{\citenamefont {Si}\ \emph {et~al.}(2015)\citenamefont {Si},
  \citenamefont {Zhong}, \citenamefont {Tomczak},\ and\ \citenamefont
  {Held}}]{Si2015}%
  \BibitemOpen
  \bibfield  {author} {\bibinfo {author} {\bibfnamefont {L.}~\bibnamefont
  {Si}}, \bibinfo {author} {\bibfnamefont {Z.}~\bibnamefont {Zhong}}, \bibinfo
  {author} {\bibfnamefont {J.~M.}\ \bibnamefont {Tomczak}}, \ and\ \bibinfo
  {author} {\bibfnamefont {K.}~\bibnamefont {Held}},\ }\href {\doibase
  10.1103/PhysRevB.92.041108} {\bibfield  {journal} {\bibinfo  {journal} {Phys.
  Rev. B}\ }\textbf {\bibinfo {volume} {92}},\ \bibinfo {pages} {041108}
  (\bibinfo {year} {2015})}\BibitemShut {NoStop}%
\bibitem [{\citenamefont {Chang}\ \emph {et~al.}(2009)\citenamefont {Chang},
  \citenamefont {Kim}, \citenamefont {Phark}, \citenamefont {Kim},
  \citenamefont {Yu},\ and\ \citenamefont {Noh}}]{Chang2009}%
  \BibitemOpen
  \bibfield  {author} {\bibinfo {author} {\bibfnamefont {Y.~J.}\ \bibnamefont
  {Chang}}, \bibinfo {author} {\bibfnamefont {C.~H.}\ \bibnamefont {Kim}},
  \bibinfo {author} {\bibfnamefont {S.-H.}\ \bibnamefont {Phark}}, \bibinfo
  {author} {\bibfnamefont {Y.~S.}\ \bibnamefont {Kim}}, \bibinfo {author}
  {\bibfnamefont {J.}~\bibnamefont {Yu}}, \ and\ \bibinfo {author}
  {\bibfnamefont {T.~W.}\ \bibnamefont {Noh}},\ }\href {\doibase
  10.1103/PhysRevLett.103.057201} {\bibfield  {journal} {\bibinfo  {journal}
  {Phys. Rev. Lett.}\ }\textbf {\bibinfo {volume} {103}},\ \bibinfo {pages}
  {057201} (\bibinfo {year} {2009})}\BibitemShut {NoStop}%
\bibitem [{\citenamefont {Kim}\ \emph {et~al.}(2017)\citenamefont {Kim},
  \citenamefont {Khmelevskyi}, \citenamefont {Mazin}, \citenamefont
  {Agterberg},\ and\ \citenamefont {Franchini}}]{BKim2017}%
  \BibitemOpen
  \bibfield  {author} {\bibinfo {author} {\bibfnamefont {B.}~\bibnamefont
  {Kim}}, \bibinfo {author} {\bibfnamefont {S.}~\bibnamefont {Khmelevskyi}},
  \bibinfo {author} {\bibfnamefont {I.~I.}\ \bibnamefont {Mazin}}, \bibinfo
  {author} {\bibfnamefont {D.~F.}\ \bibnamefont {Agterberg}}, \ and\ \bibinfo
  {author} {\bibfnamefont {C.}~\bibnamefont {Franchini}},\ }\href
  {https://www.nature.com/articles/s41535-017-0041-8} {\bibfield  {journal}
  {\bibinfo  {journal} {npj Quantum Materials}\ }\textbf {\bibinfo {volume}
  {2}},\ \bibinfo {pages} {37} (\bibinfo {year} {2017})}\BibitemShut {NoStop}%
\bibitem [{\citenamefont {Larson}\ \emph {et~al.}(2004)\citenamefont {Larson},
  \citenamefont {Mazin},\ and\ \citenamefont {Singh}}]{Larson2004}%
  \BibitemOpen
  \bibfield  {author} {\bibinfo {author} {\bibfnamefont {P.}~\bibnamefont
  {Larson}}, \bibinfo {author} {\bibfnamefont {I.~I.}\ \bibnamefont {Mazin}}, \
  and\ \bibinfo {author} {\bibfnamefont {D.~J.}\ \bibnamefont {Singh}},\ }\href
  {\doibase 10.1103/PhysRevB.69.064429} {\bibfield  {journal} {\bibinfo
  {journal} {Phys. Rev. B}\ }\textbf {\bibinfo {volume} {69}},\ \bibinfo
  {pages} {064429} (\bibinfo {year} {2004})}\BibitemShut {NoStop}%
\bibitem [{\citenamefont {Ortenzi}\ \emph {et~al.}(2012)\citenamefont
  {Ortenzi}, \citenamefont {Mazin}, \citenamefont {Blaha},\ and\ \citenamefont
  {Boeri}}]{Ortenzi2012}%
  \BibitemOpen
  \bibfield  {author} {\bibinfo {author} {\bibfnamefont {L.}~\bibnamefont
  {Ortenzi}}, \bibinfo {author} {\bibfnamefont {I.~I.}\ \bibnamefont {Mazin}},
  \bibinfo {author} {\bibfnamefont {P.}~\bibnamefont {Blaha}}, \ and\ \bibinfo
  {author} {\bibfnamefont {L.}~\bibnamefont {Boeri}},\ }\href {\doibase
  10.1103/PhysRevB.86.064437} {\bibfield  {journal} {\bibinfo  {journal} {Phys.
  Rev. B}\ }\textbf {\bibinfo {volume} {86}},\ \bibinfo {pages} {064437}
  (\bibinfo {year} {2012})}\BibitemShut {NoStop}%
\bibitem [{\citenamefont {Kugler}\ \emph {et~al.}(2019)\citenamefont {Kugler},
  \citenamefont {Zingl}, \citenamefont {Strand}, \citenamefont {Lee},
  \citenamefont {von Delft},\ and\ \citenamefont {Georges}}]{Kugler2019}%
  \BibitemOpen
  \bibfield  {author} {\bibinfo {author} {\bibfnamefont {F.~B.}\ \bibnamefont
  {Kugler}}, \bibinfo {author} {\bibfnamefont {M.}~\bibnamefont {Zingl}},
  \bibinfo {author} {\bibfnamefont {H.~U.~R.}\ \bibnamefont {Strand}}, \bibinfo
  {author} {\bibfnamefont {S.-S.~B.}\ \bibnamefont {Lee}}, \bibinfo {author}
  {\bibfnamefont {J.}~\bibnamefont {von Delft}}, \ and\ \bibinfo {author}
  {\bibfnamefont {A.}~\bibnamefont {Georges}},\ }\href
  {http://arxiv.org/abs/1909.02389} {\bibfield  {journal} {\bibinfo  {journal}
  {arXiv:1909.02389}\ } (\bibinfo {year} {2019})}\BibitemShut {NoStop}%
\bibitem [{\citenamefont {Sidis}\ \emph {et~al.}(1999)\citenamefont {Sidis},
  \citenamefont {Braden}, \citenamefont {Bourges}, \citenamefont {Hennion},
  \citenamefont {NishiZaki}, \citenamefont {Maeno},\ and\ \citenamefont
  {Mori}}]{Sidis1999}%
  \BibitemOpen
  \bibfield  {author} {\bibinfo {author} {\bibfnamefont {Y.}~\bibnamefont
  {Sidis}}, \bibinfo {author} {\bibfnamefont {M.}~\bibnamefont {Braden}},
  \bibinfo {author} {\bibfnamefont {P.}~\bibnamefont {Bourges}}, \bibinfo
  {author} {\bibfnamefont {B.}~\bibnamefont {Hennion}}, \bibinfo {author}
  {\bibfnamefont {S.}~\bibnamefont {NishiZaki}}, \bibinfo {author}
  {\bibfnamefont {Y.}~\bibnamefont {Maeno}}, \ and\ \bibinfo {author}
  {\bibfnamefont {Y.}~\bibnamefont {Mori}},\ }\href {\doibase
  10.1103/PhysRevLett.83.3320} {\bibfield  {journal} {\bibinfo  {journal}
  {Phys. Rev. Lett.}\ }\textbf {\bibinfo {volume} {83}},\ \bibinfo {pages}
  {3320} (\bibinfo {year} {1999})}\BibitemShut {NoStop}%
\bibitem [{\citenamefont {Iida}\ \emph {et~al.}(2011)\citenamefont {Iida},
  \citenamefont {Kofu}, \citenamefont {Katayama}, \citenamefont {Lee},
  \citenamefont {Kajimoto}, \citenamefont {Inamura}, \citenamefont {Nakamura},
  \citenamefont {Arai}, \citenamefont {Yoshida}, \citenamefont {Fujita},
  \citenamefont {Yamada},\ and\ \citenamefont {Lee}}]{Iida2011}%
  \BibitemOpen
  \bibfield  {author} {\bibinfo {author} {\bibfnamefont {K.}~\bibnamefont
  {Iida}}, \bibinfo {author} {\bibfnamefont {M.}~\bibnamefont {Kofu}}, \bibinfo
  {author} {\bibfnamefont {N.}~\bibnamefont {Katayama}}, \bibinfo {author}
  {\bibfnamefont {J.}~\bibnamefont {Lee}}, \bibinfo {author} {\bibfnamefont
  {R.}~\bibnamefont {Kajimoto}}, \bibinfo {author} {\bibfnamefont
  {Y.}~\bibnamefont {Inamura}}, \bibinfo {author} {\bibfnamefont
  {M.}~\bibnamefont {Nakamura}}, \bibinfo {author} {\bibfnamefont
  {M.}~\bibnamefont {Arai}}, \bibinfo {author} {\bibfnamefont {Y.}~\bibnamefont
  {Yoshida}}, \bibinfo {author} {\bibfnamefont {M.}~\bibnamefont {Fujita}},
  \bibinfo {author} {\bibfnamefont {K.}~\bibnamefont {Yamada}}, \ and\ \bibinfo
  {author} {\bibfnamefont {S.-H.}\ \bibnamefont {Lee}},\ }\href {\doibase
  10.1103/PhysRevB.84.060402} {\bibfield  {journal} {\bibinfo  {journal} {Phys.
  Rev. B}\ }\textbf {\bibinfo {volume} {84}},\ \bibinfo {pages} {060402}
  (\bibinfo {year} {2011})}\BibitemShut {NoStop}%
\bibitem [{\citenamefont {Cobo}\ \emph {et~al.}(2016)\citenamefont {Cobo},
  \citenamefont {Ahn}, \citenamefont {Eremin},\ and\ \citenamefont
  {Akbari}}]{Cobo2016}%
  \BibitemOpen
  \bibfield  {author} {\bibinfo {author} {\bibfnamefont {S.}~\bibnamefont
  {Cobo}}, \bibinfo {author} {\bibfnamefont {F.}~\bibnamefont {Ahn}}, \bibinfo
  {author} {\bibfnamefont {I.}~\bibnamefont {Eremin}}, \ and\ \bibinfo {author}
  {\bibfnamefont {A.}~\bibnamefont {Akbari}},\ }\href {\doibase
  10.1103/PhysRevB.94.224507} {\bibfield  {journal} {\bibinfo  {journal} {Phys.
  Rev. B}\ }\textbf {\bibinfo {volume} {94}},\ \bibinfo {pages} {224507}
  (\bibinfo {year} {2016})}\BibitemShut {NoStop}%
\bibitem [{\citenamefont {Mazin}\ \emph {et~al.}(2008)\citenamefont {Mazin},
  \citenamefont {Johannes}, \citenamefont {Boeri}, \citenamefont {Koepernik},\
  and\ \citenamefont {Singh}}]{Mazin2008}%
  \BibitemOpen
  \bibfield  {author} {\bibinfo {author} {\bibfnamefont {I.~I.}\ \bibnamefont
  {Mazin}}, \bibinfo {author} {\bibfnamefont {M.~D.}\ \bibnamefont {Johannes}},
  \bibinfo {author} {\bibfnamefont {L.}~\bibnamefont {Boeri}}, \bibinfo
  {author} {\bibfnamefont {K.}~\bibnamefont {Koepernik}}, \ and\ \bibinfo
  {author} {\bibfnamefont {D.~J.}\ \bibnamefont {Singh}},\ }\href {\doibase
  10.1103/PhysRevB.78.085104} {\bibfield  {journal} {\bibinfo  {journal} {Phys.
  Rev. B}\ }\textbf {\bibinfo {volume} {78}},\ \bibinfo {pages} {085104}
  (\bibinfo {year} {2008})}\BibitemShut {NoStop}%
\bibitem [{\citenamefont {Glasbrenner}\ \emph {et~al.}(2015)\citenamefont
  {Glasbrenner}, \citenamefont {Mazin}, \citenamefont {Jeschke}, \citenamefont
  {Hirschfeld}, \citenamefont {Fernandes},\ and\ \citenamefont
  {Valen\'{\i}}}]{Glasbrenner2015}%
  \BibitemOpen
  \bibfield  {author} {\bibinfo {author} {\bibfnamefont {J.~K.}\ \bibnamefont
  {Glasbrenner}}, \bibinfo {author} {\bibfnamefont {I.~I.}\ \bibnamefont
  {Mazin}}, \bibinfo {author} {\bibfnamefont {H.~O.}\ \bibnamefont {Jeschke}},
  \bibinfo {author} {\bibfnamefont {P.~J.}\ \bibnamefont {Hirschfeld}},
  \bibinfo {author} {\bibfnamefont {R.~M.}\ \bibnamefont {Fernandes}}, \ and\
  \bibinfo {author} {\bibfnamefont {R.}~\bibnamefont {Valen\'{\i}}},\
  }\href@noop {} {\bibfield  {journal} {\bibinfo  {journal} {Nature Physics}\
  }\textbf {\bibinfo {volume} {11}},\ \bibinfo {pages} {953} (\bibinfo {year}
  {2015})}\BibitemShut {NoStop}%
\bibitem [{\citenamefont {Vailionis}\ \emph {et~al.}(2011)\citenamefont
  {Vailionis}, \citenamefont {Boschker}, \citenamefont {Siemons}, \citenamefont
  {Houwman}, \citenamefont {Blank}, \citenamefont {Rijnders},\ and\
  \citenamefont {Koster}}]{Vailionis2011}%
  \BibitemOpen
  \bibfield  {author} {\bibinfo {author} {\bibfnamefont {A.}~\bibnamefont
  {Vailionis}}, \bibinfo {author} {\bibfnamefont {H.}~\bibnamefont {Boschker}},
  \bibinfo {author} {\bibfnamefont {W.}~\bibnamefont {Siemons}}, \bibinfo
  {author} {\bibfnamefont {E.~P.}\ \bibnamefont {Houwman}}, \bibinfo {author}
  {\bibfnamefont {D.~H.~A.}\ \bibnamefont {Blank}}, \bibinfo {author}
  {\bibfnamefont {G.}~\bibnamefont {Rijnders}}, \ and\ \bibinfo {author}
  {\bibfnamefont {G.}~\bibnamefont {Koster}},\ }\href {\doibase
  10.1103/PhysRevB.83.064101} {\bibfield  {journal} {\bibinfo  {journal} {Phys.
  Rev. B}\ }\textbf {\bibinfo {volume} {83}},\ \bibinfo {pages} {064101}
  (\bibinfo {year} {2011})}\BibitemShut {NoStop}%
\bibitem [{\citenamefont {Schlom}\ \emph {et~al.}(2007)\citenamefont {Schlom},
  \citenamefont {Chen}, \citenamefont {Eom}, \citenamefont {Rabe},
  \citenamefont {Streiffer},\ and\ \citenamefont {Triscone}}]{Schlom2007}%
  \BibitemOpen
  \bibfield  {author} {\bibinfo {author} {\bibfnamefont {D.~G.}\ \bibnamefont
  {Schlom}}, \bibinfo {author} {\bibfnamefont {L.-Q.}\ \bibnamefont {Chen}},
  \bibinfo {author} {\bibfnamefont {C.-B.}\ \bibnamefont {Eom}}, \bibinfo
  {author} {\bibfnamefont {K.~M.}\ \bibnamefont {Rabe}}, \bibinfo {author}
  {\bibfnamefont {S.~K.}\ \bibnamefont {Streiffer}}, \ and\ \bibinfo {author}
  {\bibfnamefont {J.-M.}\ \bibnamefont {Triscone}},\ }\href {\doibase
  10.1146/annurev.matsci.37.061206.113016} {\bibfield  {journal} {\bibinfo
  {journal} {Annu. Rev. Mater. Res.}\ }\textbf {\bibinfo {volume} {37}},\
  \bibinfo {pages} {589} (\bibinfo {year} {2007})}\BibitemShut {NoStop}%
\bibitem [{sup()}]{suppl}%
  \BibitemOpen
  \href@noop {} {\bibinfo  {journal} {See supplement materials for (i)
  calculation details, (ii) susceptibility calculation, (iii) strain-dependent
  changes of structure and DOSs, and (iv) role of spin-orbit coupling on the
  Fermi surfaces, which includes
  Refs.~\cite{Kresse1993,Kresse1996,Perdew1996,Koepernik1999,Ku2010,Barber2019,Suh2019,Shen2001,Veenstra2013}}\
  }\BibitemShut {NoStop}%
\bibitem [{com()}]{comment}%
  \BibitemOpen
\bibfield  {journal} {  }\href@noop {} {\bibinfo  {journal} {For the detailed
  structure of the DOS peak at around Fermi level, more accurate treatments are
  needed. For example, see Luo \emph{et al.}~\cite{Luo2019}. In current study,
  we focus on the general trends upon the substrate strain}\ }\BibitemShut
  {NoStop}%
\bibitem [{\citenamefont {Ko}\ \emph {et~al.}(2007)\citenamefont {Ko},
  \citenamefont {Kim}, \citenamefont {Kim},\ and\ \citenamefont
  {Choi}}]{Ko2007}%
  \BibitemOpen
\bibfield  {journal} {  }\bibfield  {author} {\bibinfo {author} {\bibfnamefont
  {E.}~\bibnamefont {Ko}}, \bibinfo {author} {\bibfnamefont {B.~J.}\
  \bibnamefont {Kim}}, \bibinfo {author} {\bibfnamefont {C.}~\bibnamefont
  {Kim}}, \ and\ \bibinfo {author} {\bibfnamefont {H.~J.}\ \bibnamefont
  {Choi}},\ }\href {\doibase 10.1103/PhysRevLett.98.226401} {\bibfield
  {journal} {\bibinfo  {journal} {Phys. Rev. Lett.}\ }\textbf {\bibinfo
  {volume} {98}},\ \bibinfo {pages} {226401} (\bibinfo {year}
  {2007})}\BibitemShut {NoStop}%
\bibitem [{\citenamefont {Pustogow}\ \emph {et~al.}(2019)\citenamefont
  {Pustogow}, \citenamefont {Luo}, \citenamefont {Chronister}, \citenamefont
  {Su}, \citenamefont {Sokolov}, \citenamefont {Jerzembeck}, \citenamefont
  {Mackenzie}, \citenamefont {Hicks}, \citenamefont {Kikugawa}, \citenamefont
  {Raghu}, \citenamefont {Bauer},\ and\ \citenamefont {Brown}}]{Pustogow2019}%
  \BibitemOpen
  \bibfield  {author} {\bibinfo {author} {\bibfnamefont {A.}~\bibnamefont
  {Pustogow}}, \bibinfo {author} {\bibfnamefont {Y.}~\bibnamefont {Luo}},
  \bibinfo {author} {\bibfnamefont {A.}~\bibnamefont {Chronister}}, \bibinfo
  {author} {\bibfnamefont {Y.-S.}\ \bibnamefont {Su}}, \bibinfo {author}
  {\bibfnamefont {D.}~\bibnamefont {Sokolov}}, \bibinfo {author} {\bibfnamefont
  {F.}~\bibnamefont {Jerzembeck}}, \bibinfo {author} {\bibfnamefont {A.~P.}\
  \bibnamefont {Mackenzie}}, \bibinfo {author} {\bibfnamefont {C.~W.}\
  \bibnamefont {Hicks}}, \bibinfo {author} {\bibfnamefont {N.}~\bibnamefont
  {Kikugawa}}, \bibinfo {author} {\bibfnamefont {S.}~\bibnamefont {Raghu}},
  \bibinfo {author} {\bibfnamefont {E.~D.}\ \bibnamefont {Bauer}}, \ and\
  \bibinfo {author} {\bibfnamefont {S.~E.}\ \bibnamefont {Brown}},\ }\href@noop
  {} {\bibfield  {journal} {\bibinfo  {journal} {Nature Physics}\ }\textbf
  {\bibinfo {volume} {574}},\ \bibinfo {pages} {72} (\bibinfo {year}
  {2019})}\BibitemShut {NoStop}%
\bibitem [{\citenamefont {Ishida}\ \emph {et~al.}(2019)\citenamefont {Ishida},
  \citenamefont {Manago},\ and\ \citenamefont {Maeno}}]{Ishida2019}%
  \BibitemOpen
  \bibfield  {author} {\bibinfo {author} {\bibfnamefont {K.}~\bibnamefont
  {Ishida}}, \bibinfo {author} {\bibfnamefont {M.}~\bibnamefont {Manago}}, \
  and\ \bibinfo {author} {\bibfnamefont {Y.}~\bibnamefont {Maeno}},\ }\href
  {http://arxiv.org/abs/1907.12236} {\bibfield  {journal} {\bibinfo  {journal}
  {arXiv:1907.12236}\ } (\bibinfo {year} {2019})}\BibitemShut {NoStop}%
\bibitem [{\citenamefont {Kresse}\ and\ \citenamefont
  {Hafner}(1993)}]{Kresse1993}%
  \BibitemOpen
  \bibfield  {author} {\bibinfo {author} {\bibfnamefont {G.}~\bibnamefont
  {Kresse}}\ and\ \bibinfo {author} {\bibfnamefont {J.}~\bibnamefont
  {Hafner}},\ }\href {\doibase 10.1103/PhysRevB.47.558} {\bibfield  {journal}
  {\bibinfo  {journal} {Phys. Rev. B}\ }\textbf {\bibinfo {volume} {47}},\
  \bibinfo {pages} {558} (\bibinfo {year} {1993})}\BibitemShut {NoStop}%
\bibitem [{\citenamefont {Kresse}\ and\ \citenamefont
  {Furthm\"uller}(1996)}]{Kresse1996}%
  \BibitemOpen
  \bibfield  {author} {\bibinfo {author} {\bibfnamefont {G.}~\bibnamefont
  {Kresse}}\ and\ \bibinfo {author} {\bibfnamefont {J.}~\bibnamefont
  {Furthm\"uller}},\ }\href {\doibase 10.1103/PhysRevB.54.11169} {\bibfield
  {journal} {\bibinfo  {journal} {Phys. Rev. B}\ }\textbf {\bibinfo {volume}
  {54}},\ \bibinfo {pages} {11169} (\bibinfo {year} {1996})}\BibitemShut
  {NoStop}%
\bibitem [{\citenamefont {Perdew}\ \emph {et~al.}(1996)\citenamefont {Perdew},
  \citenamefont {Burke},\ and\ \citenamefont {Ernzerhof}}]{Perdew1996}%
  \BibitemOpen
  \bibfield  {author} {\bibinfo {author} {\bibfnamefont {J.~P.}\ \bibnamefont
  {Perdew}}, \bibinfo {author} {\bibfnamefont {K.}~\bibnamefont {Burke}}, \
  and\ \bibinfo {author} {\bibfnamefont {M.}~\bibnamefont {Ernzerhof}},\ }\href
  {\doibase 10.1103/PhysRevLett.77.3865} {\bibfield  {journal} {\bibinfo
  {journal} {Phys. Rev. Lett.}\ }\textbf {\bibinfo {volume} {77}},\ \bibinfo
  {pages} {3865} (\bibinfo {year} {1996})}\BibitemShut {NoStop}%
\bibitem [{\citenamefont {Koepernik}\ and\ \citenamefont
  {Eschrig}(1999)}]{Koepernik1999}%
  \BibitemOpen
  \bibfield  {author} {\bibinfo {author} {\bibfnamefont {K.}~\bibnamefont
  {Koepernik}}\ and\ \bibinfo {author} {\bibfnamefont {H.}~\bibnamefont
  {Eschrig}},\ }\href {\doibase 10.1103/PhysRevB.59.1743} {\bibfield  {journal}
  {\bibinfo  {journal} {Phys. Rev. B}\ }\textbf {\bibinfo {volume} {59}},\
  \bibinfo {pages} {1743} (\bibinfo {year} {1999})}\BibitemShut {NoStop}%
\bibitem [{\citenamefont {Ku}\ \emph {et~al.}(2010)\citenamefont {Ku},
  \citenamefont {Berlijn},\ and\ \citenamefont {Lee}}]{Ku2010}%
  \BibitemOpen
  \bibfield  {author} {\bibinfo {author} {\bibfnamefont {W.}~\bibnamefont
  {Ku}}, \bibinfo {author} {\bibfnamefont {T.}~\bibnamefont {Berlijn}}, \ and\
  \bibinfo {author} {\bibfnamefont {C.-C.}\ \bibnamefont {Lee}},\ }\href
  {\doibase 10.1103/PhysRevLett.104.216401} {\bibfield  {journal} {\bibinfo
  {journal} {Phys. Rev. Lett.}\ }\textbf {\bibinfo {volume} {104}},\ \bibinfo
  {pages} {216401} (\bibinfo {year} {2010})}\BibitemShut {NoStop}%
\bibitem [{\citenamefont {Barber}\ \emph {et~al.}(2019)\citenamefont {Barber},
  \citenamefont {Lechermann}, \citenamefont {Streltsov}, \citenamefont
  {Skornyakov}, \citenamefont {Ghosh}, \citenamefont {Ramshaw}, \citenamefont
  {Kikugawa}, \citenamefont {Sokolov}, \citenamefont {Mackenzie}, \citenamefont
  {Hicks},\ and\ \citenamefont {Mazin}}]{Barber2019}%
  \BibitemOpen
  \bibfield  {author} {\bibinfo {author} {\bibfnamefont {M.~E.}\ \bibnamefont
  {Barber}}, \bibinfo {author} {\bibfnamefont {F.}~\bibnamefont {Lechermann}},
  \bibinfo {author} {\bibfnamefont {S.~V.}\ \bibnamefont {Streltsov}}, \bibinfo
  {author} {\bibfnamefont {S.~L.}\ \bibnamefont {Skornyakov}}, \bibinfo
  {author} {\bibfnamefont {S.}~\bibnamefont {Ghosh}}, \bibinfo {author}
  {\bibfnamefont {B.~J.}\ \bibnamefont {Ramshaw}}, \bibinfo {author}
  {\bibfnamefont {N.}~\bibnamefont {Kikugawa}}, \bibinfo {author}
  {\bibfnamefont {D.~A.}\ \bibnamefont {Sokolov}}, \bibinfo {author}
  {\bibfnamefont {A.~P.}\ \bibnamefont {Mackenzie}}, \bibinfo {author}
  {\bibfnamefont {C.~W.}\ \bibnamefont {Hicks}}, \ and\ \bibinfo {author}
  {\bibfnamefont {I.~I.}\ \bibnamefont {Mazin}},\ }\href {\doibase
  10.1103/PhysRevB.100.245139} {\bibfield  {journal} {\bibinfo  {journal}
  {Phys. Rev. B}\ }\textbf {\bibinfo {volume} {100}},\ \bibinfo {pages}
  {245139} (\bibinfo {year} {2019})}\BibitemShut {NoStop}%
\bibitem [{\citenamefont {Suh}\ \emph {et~al.}(2019)\citenamefont {Suh},
  \citenamefont {Menke}, \citenamefont {Brydon}, \citenamefont {Timm},
  \citenamefont {Ramires},\ and\ \citenamefont {Agterberg}}]{Suh2019}%
  \BibitemOpen
  \bibfield  {author} {\bibinfo {author} {\bibfnamefont {H.~G.}\ \bibnamefont
  {Suh}}, \bibinfo {author} {\bibfnamefont {H.}~\bibnamefont {Menke}}, \bibinfo
  {author} {\bibfnamefont {P.}~\bibnamefont {Brydon}}, \bibinfo {author}
  {\bibfnamefont {C.}~\bibnamefont {Timm}}, \bibinfo {author} {\bibfnamefont
  {A.}~\bibnamefont {Ramires}}, \ and\ \bibinfo {author} {\bibfnamefont
  {D.~F.}\ \bibnamefont {Agterberg}},\ }\href {http://arxiv.org/abs/1912.09525}
  {\bibfield  {journal} {\bibinfo  {journal} {arXiv:1912.09525}\ } (\bibinfo
  {year} {2019})}\BibitemShut {NoStop}%
\bibitem [{\citenamefont {Shen}\ \emph {et~al.}(2001)\citenamefont {Shen},
  \citenamefont {Damascelli}, \citenamefont {Lu}, \citenamefont {Armitage},
  \citenamefont {Ronning}, \citenamefont {Feng}, \citenamefont {Kim},
  \citenamefont {Shen}, \citenamefont {Singh}, \citenamefont {Mazin},
  \citenamefont {Nakatsuji}, \citenamefont {Mao}, \citenamefont {Maeno},
  \citenamefont {Kimura},\ and\ \citenamefont {Tokura}}]{Shen2001}%
  \BibitemOpen
  \bibfield  {author} {\bibinfo {author} {\bibfnamefont {K.~M.}\ \bibnamefont
  {Shen}}, \bibinfo {author} {\bibfnamefont {A.}~\bibnamefont {Damascelli}},
  \bibinfo {author} {\bibfnamefont {D.~H.}\ \bibnamefont {Lu}}, \bibinfo
  {author} {\bibfnamefont {N.~P.}\ \bibnamefont {Armitage}}, \bibinfo {author}
  {\bibfnamefont {F.}~\bibnamefont {Ronning}}, \bibinfo {author} {\bibfnamefont
  {D.~L.}\ \bibnamefont {Feng}}, \bibinfo {author} {\bibfnamefont
  {C.}~\bibnamefont {Kim}}, \bibinfo {author} {\bibfnamefont {Z.-X.}\
  \bibnamefont {Shen}}, \bibinfo {author} {\bibfnamefont {D.~J.}\ \bibnamefont
  {Singh}}, \bibinfo {author} {\bibfnamefont {I.~I.}\ \bibnamefont {Mazin}},
  \bibinfo {author} {\bibfnamefont {S.}~\bibnamefont {Nakatsuji}}, \bibinfo
  {author} {\bibfnamefont {Z.~Q.}\ \bibnamefont {Mao}}, \bibinfo {author}
  {\bibfnamefont {Y.}~\bibnamefont {Maeno}}, \bibinfo {author} {\bibfnamefont
  {T.}~\bibnamefont {Kimura}}, \ and\ \bibinfo {author} {\bibfnamefont
  {Y.}~\bibnamefont {Tokura}},\ }\href {\doibase 10.1103/PhysRevB.64.180502}
  {\bibfield  {journal} {\bibinfo  {journal} {Phys. Rev. B}\ }\textbf {\bibinfo
  {volume} {64}},\ \bibinfo {pages} {180502} (\bibinfo {year}
  {2001})}\BibitemShut {NoStop}%
\bibitem [{\citenamefont {Veenstra}\ \emph {et~al.}(2013)\citenamefont
  {Veenstra}, \citenamefont {Zhu}, \citenamefont {Ludbrook}, \citenamefont
  {Capsoni}, \citenamefont {Levy}, \citenamefont {Nicolaou}, \citenamefont
  {Rosen}, \citenamefont {Comin}, \citenamefont {Kittaka}, \citenamefont
  {Maeno}, \citenamefont {Elfimov},\ and\ \citenamefont
  {Damascelli}}]{Veenstra2013}%
  \BibitemOpen
  \bibfield  {author} {\bibinfo {author} {\bibfnamefont {C.~N.}\ \bibnamefont
  {Veenstra}}, \bibinfo {author} {\bibfnamefont {Z.-H.}\ \bibnamefont {Zhu}},
  \bibinfo {author} {\bibfnamefont {B.}~\bibnamefont {Ludbrook}}, \bibinfo
  {author} {\bibfnamefont {M.}~\bibnamefont {Capsoni}}, \bibinfo {author}
  {\bibfnamefont {G.}~\bibnamefont {Levy}}, \bibinfo {author} {\bibfnamefont
  {A.}~\bibnamefont {Nicolaou}}, \bibinfo {author} {\bibfnamefont {J.~A.}\
  \bibnamefont {Rosen}}, \bibinfo {author} {\bibfnamefont {R.}~\bibnamefont
  {Comin}}, \bibinfo {author} {\bibfnamefont {S.}~\bibnamefont {Kittaka}},
  \bibinfo {author} {\bibfnamefont {Y.}~\bibnamefont {Maeno}}, \bibinfo
  {author} {\bibfnamefont {I.~S.}\ \bibnamefont {Elfimov}}, \ and\ \bibinfo
  {author} {\bibfnamefont {A.}~\bibnamefont {Damascelli}},\ }\href {\doibase
  10.1103/PhysRevLett.110.097004} {\bibfield  {journal} {\bibinfo  {journal}
  {Phys. Rev. Lett.}\ }\textbf {\bibinfo {volume} {110}},\ \bibinfo {pages}
  {097004} (\bibinfo {year} {2013})}\BibitemShut {NoStop}%
\end{thebibliography}%

\section{Acknowledgements}
We thank Beom Hyun Kim and Minjae Kim for fruitful discussions.
This work was supported by the NRF Grant (Contracts No. 2018R1D1A1A02086051
and 2016R1D1A1B02008461), the research funds of Kunsan National University,
and Max-Plank POSTECH/KOREA Research Initiative (Grant No.
2016K1A4A4A01922028). The computing resources from the KISTI supercomputing
center (Project No. KSC-2018-CRE-0079) is greatly acknowledged. I.I.M.
acknowledges support by ONR through the NRL basic research program.


\vspace*{250px} \newpage


\renewcommand{\thefigure}{S\arabic{figure}}
\renewcommand{\thetable}{S\arabic{table}}
\renewcommand{\bibnumfmt}[1]{[S#1]}
\setcounter{table}{0}
\setcounter{figure}{0}
\setcounter{equation}{0}

\onecolumngrid

\clearpage

\begin{center}
{\bf \Large
\textit{Supplementary Material:}\\SrRuO$_3$-SrTiO$_3$ heterostructure as a possible platform for studying unconventional superconductivity in  Sr$_{2}$RuO$_{4}$}

\vspace{0.2 cm}

{\large
Bongjae Kim,$^{1,2}$ Sergii Khmelevskyi,$^{3}$ Cesare Franchini,$^{4,5}$ I. I. Mazin,$^{6,7}$ and Kyoo Kim$^{2,8,9}$
}

\vspace{0.1 cm}

{\it
$^{1}$ Department of Physics, Kunsan National University, Gunsan, 54150, Korea

$^{2}$ MPPHC-CPM, Max Planck POSTECH/Korea Research Initiative, Pohang 37673, Korea

$^{3}$ Center for Computational Materials Science, Institute for Applied
Physics, Vienna University of Technology, Wiedner Hauptstrasse $8$ - $10$,
$1040$ Vienna, Austria

$^{4}$ University of Vienna, Faculty of Physics and Center for Computational
Materials Science, Vienna A-1090, Austria

$^{5}$ Dipartimento di Fisica e Astronomia, Universit\`{a} di Bologna, 40127
Bologna, Italy

$^{6}$ Code 6393, Naval Research Laboratory, Washington, DC 20375, USA

$^{7}$ Quantum Materials Center, George Mason
University, Fairfax, VA 22030, USA

$^{8}$ Department of Physics, Pohang University of Science and Technology,
Pohang 37673, Korea

$^{9}$ Korea Atomic Energy Research Institute (KAERI), 111 Daedeok-daero, Daejeon 34057, Korea
}

\end{center}

\section{Calculation details}

We performed \textit{ab initio} electronic structure calculations using
projector augmented wave method employing the Vienna \textit{ab initio}
simulation package (VASP) \cite{Kresse1993,Kresse1996}, the generalized
gradient approximation by Perdew-Burke-Ernzerhof with plane-wave cutoff of 400
eV~\cite{Perdew1996}. For SRO-STO superlattice, we fixed in-plane lattice
parameters to those of STO substrate, and varied strain up to 4\% for tensile
and compressive cases. Full atomic relaxation is performed for all cases, and
Monkhorst-Pack $k$-mesh is used as corresponding to Ref.~\cite{BKim2017}. We
also double checked the results with full potential local orbitals (FPLO)
package~\cite{Koepernik1999} For the unfolding scheme and susceptibility
calculation, we have utilized FPLO code.

\section{Susceptibility calculation}

To compare the noninteracting susceptibility of SRO-STO with that of Sr$_{2}%
$RuO$_{4}$ on equal footing, we adopted the band-unfolding scheme of
Ref.~\cite{Ku2010} to take into account of unit cell doubling induced by the
octahedral rotations. The bare susceptibility of unfolded band for Ru
$4d_{yz/zx}$ orbitals, a source of SDW fluctuation, is calculated as
\begin{equation}
Re\chi(q,\omega)=\sum_{a,b,b^{\prime},k}\frac{A^{ab}(k,\omega)A^{ab^{\prime}%
}(k+q,\omega)f^{b}(k)(1-f^{b^{\prime}}(k+q))}{E_{k}^{b}-E_{k+q}^{b^{\prime}%
}-\omega-i\delta},\nonumber
\end{equation}
where $A^{ab}(k)$ is the unfolded spectral weight, $a$ corresponds to the
bands in the unit cell without rotation (similar to that of Sr$_{2}$RuO$_{4}%
)$. Here, $4d_{xy}$ and $4d_{xz}$ orbitals contribute to quasi-1D bands, and
we only considered these bands, $a=\alpha,\beta$. $b$ and $b^{\prime}$ are
band indices of the supercell corresponding to the SRO-STO unit cell. When
there is no octahedral distortion, the spectral weights for shadow bands
vanish and the above equation describes the usual Lindhard susceptibility.

\section{Strain-dependent DOS and structural parameters}

\begin{figure}[h]
\begin{center}
\includegraphics[angle=270,width=180mm]{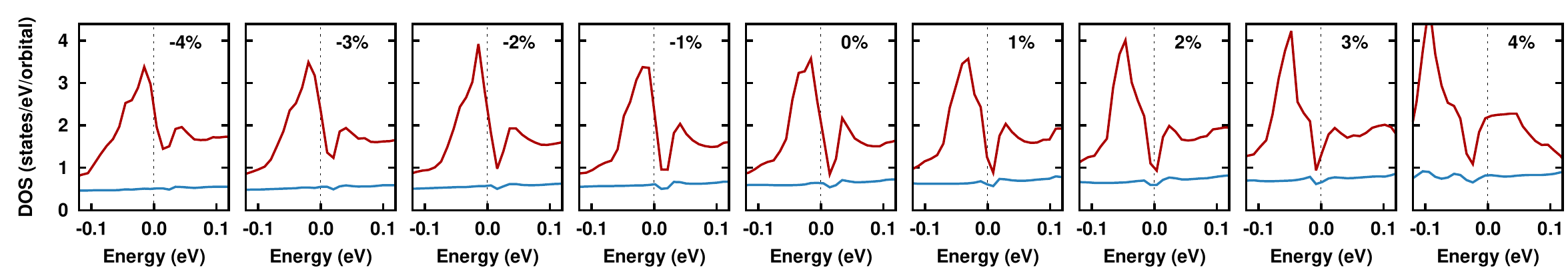}
\end{center}
\caption{ The strain-dependent DOS evolution of $xy$ (red) and $yz/zx$
orbitals at around the Fermi level.}%
\label{dos}%
\end{figure}

The strain dependence of the DOSs is shown in Fig.~\ref{dos}. As the strain
evolves from compressive to tensile, the DOS at Fermi level, $N(E_{F}$),
changes strongly.
The relative position of the van Hove peak barely changes
for compressive strain ranging from -4\% to 0\%, but progressively moves away from the $E_{F}$
for tensile strain regime.
One should note that for a better description, a more exact
treatment, as described in Ref.~\cite{Luo2019} is needed, which takes into
account renormalization due to spin fluctuations. Also, electronic correlation
and spin-orbit coupling will change some details of the van Hove
peak\cite{Barber2019}, which are however not important for our paper.

\begin{table}[th]
\caption{Ru-O-Ru bond angle ($\alpha$) and in-plane Ru-O bond length ($d_{IP}$) for
different strains. Angles are in degrees ($^{\circ}$) and length scales are in
({\AA })}%
\label{bond}
\begin{ruledtabular}
\begin{tabular}{c|ccccccccc}
strain   &    -4\%  &    -3\%  &    -2\%    &    -1\%    &    0\%    &    1\%&    2\%&    3\%&    4\%      \\
\hline
$\alpha$ &   11.3   &  10.4    &    9.5    &     8.7   &     8.2   &    8.1&    7.7&    7.3&    8.5      \\
$d_{IP}$ &   1.91   &  1.93    &   1.94    &   1.96    &   1.97    &   1.99&   2.01&   2.03&   2.05      \\
\end{tabular}
\end{ruledtabular}
\end{table}

\section{Effect of spin-orbit coupling on the Fermi surface}

\begin{figure}[h]
\begin{center}
\includegraphics[angle=0,width=100mm]{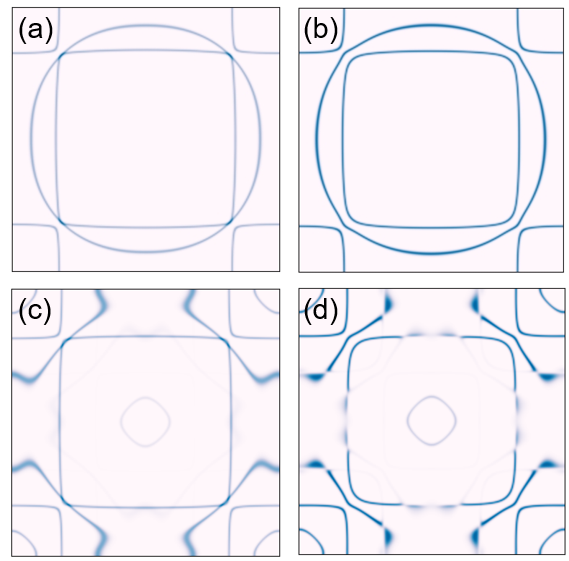}
\end{center}
\caption{The FSs of SRO-STO in the absence of octahedral rotation (a) without and (b) with the consideration of spin-orbit coupling.
(c) and (d) are unfolded FSs with and without spin-orbit coupling, respectively, when the octahedral rotation is present. }%
\label{fs}%
\end{figure}

For SRO-STO, the overall effect of spin-orbit coupling on the FS is very similar to the case of Sr$_{2}$RuO$_{4}$.
In Fig.~\ref{fs}, we have compared the FSs with and without spin-orbit coupling. For the case without the octahedral rotation,
the FSs show very similar characteristics of those from Sr$_{2}$RuO$_{4}$ (See Fig.~\ref{fs}(a) and (b)).
The spin-orbit coupling does not change the overall shape of the FSs but shows clear splitting at the crossing point between
$\beta$ and $\gamma$ sheets. This behavior is also observed with the inclusion of octahedral rotation (Fig.~\ref{fs}(c) and (d)).
Despite the changes in the FSs are not spectacular, spin-orbit coupling can play an important role in deciding the superconducting
ground states for Sr$_{2}$RuO$_{4}$~\cite{Suh2019}, and we believe that can be naturally applied to the case of SRO-STO.
The effect of octahedral rotation and subsequent hybridization of the $xy$ and $x^{2}-y^{2}$ bands are reported for the surface of
Sr$_{2}$RuO$_{4}$~\cite{Shen2001,Veenstra2013}. Similar effects are essentially observed in SRO-STO case, other than the details of the band morphology and occupations.

\end{document}